\documentclass[onecolumn,12pt]{revtex4-1}

\usepackage[english]{babel}
\usepackage{color}
\usepackage{amsmath}
\usepackage{physics}
\usepackage[hidelinks]{hyperref}
\usepackage{diagbox}
\usepackage{multirow}
\usepackage{amsfonts}
\usepackage{amssymb}
\usepackage{latexsym}
\usepackage{graphicx}
\usepackage{adjustbox}
\usepackage{float}
\usepackage{mathrsfs}
\usepackage{subcaption}
\usepackage{caption}

\newcommand{\al}{\alpha}

\newcommand{\pa}{\partial}

\newcommand{\veps}{\varepsilon}

\newcommand{\la}{\lambda}

\newcommand{\Om}{\Omega}

\newcommand{\de}{\delta}

\newcommand{\De}{\Delta}

\newcommand{\rar}{\rightarrow}

\newcommand{\non}{\nonumber}

\begin{document}

\title{Radial Anharmonic Oscillator: Perturbation Theory, New Semiclassical Expansion, Approximating Eigenfunctions.\\ I. Generalities, Cubic Anharmonicity Case}

\author{J. C. del Valle}
\email{delvalle@correo.nucleares.unam.mx}

\author{A. V. Turbiner}
\email{turbiner@nucleares.unam.mx, alexander.turbiner@stonybrook.edu}
\affiliation{Instituto de Ciencias Nucleares, Universidad Nacional Aut\'onoma de M\'exico, A. Postal 70-543 C. P. 04510, Ciudad de M\'exico, M\'exico.}

\begin{abstract}
{\small
For the general $D$-dimensional radial anharmonic oscillator with potential
$V(r)= \frac{1}{g^2}\,\hat{V}(gr)$ the Perturbation Theory (PT) in powers of coupling constant
$g$ (weak coupling regime) and in inverse, fractional powers of $g$ (strong coupling regime)
is developed constructively in $r$-space and in $(gr)$ space, respectively.
The Riccati-Bloch (RB) equation and Generalized Bloch (GB) equation are introduced as ones
which govern dynamics in coordinate $r$-space and in $(gr)$-space, respectively, exploring the logarithmic derivative of wavefunction $y$.
It is shown that PT in powers of $g$ developed in RB equation leads to Taylor expansion of $y$ at small $r$ while being developed in GB equation leads to a new form of semiclassical expansion at large $(g r)$: it coincides with loop expansion in path integral formalism. In complementary way PT for large $g$ developed in RB equation leads to an expansion of $y$ at large $r$ and developed in GB equation leads to an expansion at small $(g r)$. Interpolating all four expansions for $y$ leads to a compact function (called the {\it Approximant}), which should uniformly approximate the exact eigenfunction at $r \in [0, \infty)$ for any coupling constant $g \geq 0$ and dimension $D > 0$. 3 free parameters of the Approximant are fixed by taking it as a trial function in variational calculus. As a concrete application the low-lying states of the cubic anharmonic oscillator $V=r^2+gr^3$ are considered. It is shown that the relative deviation of the Approximant from the exact ground state eigenfunction is $\lesssim 10^{-4}$ for $r \in [0, \infty)$ for coupling constant $g \geq 0$ and dimension $D=1,2,\ldots$. In turn, the variational energies of the low-lying states are obtained with unprecedented accuracy 7-8 s.d. for $g \geq 0$ and $D=1,2,\ldots$.}
\end{abstract}

\maketitle

\section*{Introduction}

The Hamiltonian of the $D$-dimensional radial anharmonic oscillator
is given by
\begin{equation}
\hat{H}\ =\ -\frac{\hbar^2}{2M}\sum_{k=1}^{D}\pa_{ x_k}^2\ +\ V(r)\quad ,
\quad V(r)\ =\ \frac{1}{g^2}\,\hat{V}(gr)\ \equiv \ \frac{1}{g^2}\sum\limits_{k=2}^{m} a_k\, g^k\, r^k\ ,
\label{Hamiltonian}
\end{equation}
where $D=1$, $2,$ $...$,\ $r=\left(\sum_{k=1}^{D}x_k^2\right)^{1/2}$ is the radius and $M$ is the mass of the particle, $\pa_{ x_k} \equiv \frac{\pa}{\pa {x_k}}$.  Choosing the units in such a way that $\hbar=1$ and $M=1$ the coupling constant $g \geq 0$ is of dimension $[cm]^{-1}$, $a_k, k=2,\ldots m$ are real dimensionless parameters; it is assumed that $a_2, a_m > 0$, hence, $V(r)$ has minimum at $r=0$. At $m > 3$ the potential can have additional minima. For the sake of simplicity, we assume that none of them is degenerate with one at $r=0$ and the potential is positive, $V(r)>0$ at $r>0$, thus, the minimum at $r=0$ is global.
Evidently, the Hamiltonian (\ref{Hamiltonian}) has the infinite discrete spectra.
Needless to say, its study is of fundamental interest in different branches of physics, it plays an important role in atomic-molecular physics, plasma, nuclear and solid state physics. In particular, at $D=3$ with a proper choice of parameters $\{a_k\}$ and relaxing the requirement that at $r=0$ there is a global minimum, (\ref{Hamiltonian}) can be used to describe the low-lying ro-vibrational states of a diatomic molecule. Also, the Hamiltonian of the anharmonic oscillator represents a scalar quantum field theory in $(0+1)$ dimensions with an associated Lagrangian of the form
\begin{equation}
\label{LFT}
 L\ =\ \frac{1}{2}\,\dot{\vec{\varphi}}\ -\ \frac{1}{g^2}\,\sum_{k=2}^{m} \, a_k \, g^k(\vec{\varphi}\,^2)^{k/2}   \ ,
\end{equation}
where $\vec{\varphi}=(\varphi_1,...,\varphi_D)$ is a $D$-dimensional vector field.  Many properties that appear in (0+1) dimensions like asymptotic perturbation theory can also be present in a realistic field theory \cite{BENDER}. It is well known that quartic field theory at $m=4$ is renormalizable in $(3+1)$ dimensions, see e.g. \cite{LIPATOV}, while cubic ($m=3$) and sextic ($m=6$) ones are renormalizable in six and three dimensions, respectively.

There were published many papers devoted to exploration of the spectra of the Hamiltonian (\ref{Hamiltonian}) for some concrete potentials at $m=4,6$. It is worth mentioning several methods, which have been used in the past: Variational method \cite{Turbiner2005,Turbiner2010}, Rayleigh-Ritz method \cite{Taseli2dr4}, Perturbation Theory (PT) in both the weak and strong coupling regimes  \cite{BENDER,BENDERII,WENIGER,GRAFFI,SILVERSTONE,ELETSKY,FERNANDEZ},
Pad\'e Approximants \cite{LOEFFEL}, Hill Determinant method \cite{WITWIT2dr4,BISWAS,Tater},
WKB method \cite{Lu,HIOE1975}, Self-Similar Approximations \cite{YUKALOVA}, the method of Characteristic Functions \cite{ARI} among many others.

The present paper aims primarily to study the ground state of the Hamiltonian (\ref{Hamiltonian})
using as tools PT in both the weak and strong coupling regimes, the asymptotic series expansions of the eigenfunction at small and large distances and a \textit{new version} of the semiclassical approximation. Our ultimate goal is to construct (for any $r \in [0,\infty)$) a locally-accurate approximation of the ground state wave function in arbitrary (integer) dimension $D > 0$. We call this approximation the \textit{Approximant} and it is denoted by $\Psi_{0,0}^{(t)}$. Moreover, we will show how the Approximant $\Psi_{0,0}^{(t)}$  can be modified to construct Approximants for the wave functions of excited states.
It will be demonstrated that any of the above Approximants being used as trial function in variational calculation leads to highly accurate variational energy. Furthermore, if the trial function is used as the zeroth order approximation in PT scheme, a fastly-convergent PT occurs for both energy and wave function.
A similar program was implemented  in \cite{Turbiner2005,Turbiner2010} for the low-lying states of the one-dimensional  quartic anharmonic oscillator leading to an extremely high accuracy for the energy: $\sim 10$ correct significant digits (s.d.) for any $g \geq 0$ and also a locally accurate wave function (not less than 6 decimal digits (d.d.) of relative accuracy for any $g\geq 0$ and $r \in [0,\infty)$). In fact, the approximation of the wave function given in \cite{Turbiner2005,Turbiner2010} can be regarded as a {\it solution} of a non-solvable, one-dimensional quartic anharmonic oscillator. It will be shown in this paper that a similar situation occurs for the $D$-dimensional cubic radial anharmonic potential.

Let us take the ground state wave function $\Psi$ in the exponential representation
\begin{equation}
               \Psi\ =\ e^{-\frac{1}{\hbar}\Phi}\ ,
\label{exponentialrep}
\end{equation}
and call $\Phi$ the \textit{phase} of the wave function $\Psi$. It is well known in one-dimensional case that the exponential representation (\ref{exponentialrep}) being substituted into the Schr\"odinger equation leads to the nonlinear Riccati equation for the derivative of the phase \cite{LANDAU},
\begin{equation}
           \hbar\,y' \ -\  y^2\ =\ 2M(E\ -\  V)\ ,\quad y=\Phi'\ ,
\label{riccati1d}
\end{equation}
where $E$ is the energy and $V$ is the potential. Similar equation appears for the $D$-dimensional radial case.
The Riccati equation (sometimes also called the Bloch equation), by choosing the zero order approximation $y_0$ accordingly, can be solved iteratively keeping $E$ fixed, by considering the expansion in powers of Planck constant,
\begin{equation}
    y\ =\ y_0\ +\ \hbar y_1\ +\ \hbar^2 y_2\ +\ \ldots \
\end{equation}
looking for corrections $y_n$, $ n=0,1,2,...$\ . Such an iterative procedure leads to the well-known semiclassical WKB approximation, see e.g. \cite{LANDAU}. It is well-known that this expansion has zero radius of convergence, since $\hbar$ stands in front of leading derivative in (\ref{riccati1d}).

Another iterative approach is represented by the so-called Non-Linearization Procedure \cite{TURBINER:1984}. Assuming a potential of the form
\begin{equation}
      V\ =\ V_0\ +\ \la\,V_1\ +\ \la^2\,V_2\ +\ \ldots\ ,
\end{equation}
we construct PT looking for both $y$ and $E$ in the form of Taylor series in powers of the (formal) parameter $\la$. Evidently, this approach can be straightforwardly generalized to radial potentials in arbitrary dimension $D$.  It is well known that the Non-Linearization Procedure seems highly appropriate to study the bound states for the general potential in (\ref{Hamiltonian}).

For the one-dimensional anharmonic oscillator, a novel iterative approach was recently discover based on
transforming the Riccati equation (\ref{riccati1d}) into GB Equation, see  \cite{ESCOBARI,ESCOBARII,Shuryak}. It turned out this approach is also appropriate for the construction of a type of semiclassical expansion for path integral in Euclidian time for the density matrix in coinciding points with respect to quantum and thermal fluctuations around classical path (flucton).
Later it will be revealed a certain connection between the standard semiclassical WKB approximation, the Non-Linearization Procedure and the PT for the GB equation. This connection, which seems already highly non-trivial even for one-dimensional case \cite{Shuryak}, will be established for the $D$-dimensional radial anharmonic oscillator (\ref{Hamiltonian}). Note that the interpretation of a new semiclassical expansion in terms of the loop expansion for path integral in Euclidian time is not yet clear to the present authors and
it waits to be uncovered.
We discuss and exploit the above connection in order to design a prescription for construction of the Approximant $\Psi_{0,0}^{(t)}$ - the function which provides uniform approximation of the ground state function in $r, g, D$.

As illustration we present detailed results for the cubic anharmonic oscillator, seemingly for the first time. Based on the Non-Linearization Procedure we study the structure of the perturbative corrections when PT in powers of $g$ is developed. Using the Approximant as a trial function, we perform numerical variational calculations to compute the energy (and estimate its accuracy) for some low-lying states in different $D$.
The deviation of the Approximant from the exact ground state function is also estimated in framework of Non-Linearization Procedure. In addition, we calculate the first two dominant terms in the strong coupling expansion of the ground state energy using the Approximant.

The paper is organized in the following way. In the Section \ref{gen} starting from
the Riccati equation for
logarithmic derivative of the wave function the RB and the GR equations are introduced.
Section \ref{Weak Coupling} is dedicated to study the domain of weak coupling regime based on RB and GR equations. The strong coupling regime is explored in Section \ref{Strong Coupling}, in Section \ref{Approximant} the uniform approximation of the phase and ground state eigenfunction is built. The cubic anharmonic oscillator is described in details in Section \ref{cubicAHO}.

\numberwithin{equation}{section}
\section{Generalities}
\label{gen}

The Schr\"odinger equation for  a spherical-symmetric potential $V(r)$ in a $D$-dimensional space has the form
\begin{equation}
    \left[-\frac{\hbar^2}{2M}\nabla^2_D\ +\ V(r)\right]\psi\ =\ E\,\psi\ ,
\label{SE}
\end{equation}
where  $D=1,2, \ldots $ and $\nabla^2_D=\sum_{k=1}^{D}\pa^2_{x_k}$ is the $D$-dimensional Laplacian. Following the spherical symmetry of the potential, we introduce in (\ref{SE}) the $D$-dimensional hyperspherical coordinates $\{r,\Om\}$, where $r$ is the hyperradius and $\Om$ is solid angle parametrized
by $(D-1)$ Euler angles. In these coordinates the Laplacian $\nabla^2_D$ takes the form
\begin{equation}
    \nabla^2_D\ =\ \pa_r^2\ +\ \frac{D-1}{r}\pa_r\ +\ \frac{\De_{S^{D-1}}}{r^2}
    \ ,\quad  \pa_r \equiv \frac{\pa}{\pa r}\ ,
\label{laplacian}
\end{equation}
where $\De_{S^{D-1}}$ is the Laplacian on the $S^{D-1}$ sphere.
Indeed, $\De_{S^{D-1}}\, =\, -\hat{\mathbf{L}}^2$, where $\hat{\mathbf{L}}$ is the $D$-dimensional
angular momentum operator. Two remarks in row: $(i)$ for any spherically symmetric potential the $D$-dimensional angular momentum $\hat{\textbf{L}}$ is conserved, $(ii)$  in hyperspherical coordinates we can separate the hyperradial coordinate $r$ from the angular coordinates $\Om$. Hence, in $D>1$ the wave function $\psi$ can be labeled by radial quantum number $n_r$, angular quantum momentum $\ell$ and $(D-2)$ magnetic quantum numbers, it is represented as the product of two functions
\begin{equation}
\label{separation}
  \psi_{n_r,\ell,\{m_{\ell}\}}(r,\Om)\ =\ \Psi_{n_r,\ell}(r)\,\xi_{\ell,\{m_{\ell}\}}(\Om)\ ,
\end{equation}
where  $n_r=0,1,2, \ldots $\,, $\ell=0,1,2, ...$ . The set of $(D-2)$ magnetic quantum numbers $\{m_{\ell}\}$, where each of them takes values from $-\ell$ to $+\ell$, in total contains
$\mathcal{N}(D,\ell)$ different configurations, where
\begin{equation}
     \mathcal{N}(D,\ell)\ =\ \frac{(2\ell+D-2)\,(\ell+D-3)!}{\ell!\,(D-2)!}\ ,
\end{equation}
and $\mathcal{N}(D,0)\ =\ 1$.
The angular part of the wave function (for $D>1$) $\xi_{\ell,\{m_{\ell}\}}(\Om)$ corresponds to a $D$-dimensional spherical harmonic \cite{CLAUS,spherical}, which satisfies the eigenvalue equation
\begin{equation}
\label{sphlap}
-\Delta_{S^{D-1}}\, \xi_{\ell,\{m_{\ell}\}}\ =\ \hat{\mathbf{L}}^2\, \xi_{\ell,\{m_{\ell}\}} \ =\ \ell\,(\ell+D-2)\,\xi_{\ell,\{m_{\ell}\}}\ .
\end{equation}
Hence, for given $\ell$ this equation is satisfied by $\mathcal{N}(D,\ell)$ different orthogonal
spherical harmonics, which correspond to the same eigenvalue $\ell\,(\ell+D-2)$. Its degeneracy is equal to $\mathcal{N}(D,\ell)$. Note that for $D=3$ the degeneracy is given by the familiar formula $\mathcal{N}(3,\ell)=2\ell+1$.
Since the theory of spherical harmonics is well established, we should focus solely on determining the radial part of the wavefunction $\Psi_{n_r,\ell}(r)$.

After substituting (\ref{laplacian}), (\ref{separation}) and (\ref{sphlap}) to (\ref{SE}) we eventually arrive at the radial Schr\"odinger equation, which determine $\Psi_{n_r,\ell}(r)$,
\begin{equation}
\left[-\frac{\hbar^2}{2M}\left(\pa_r^2\ + \frac{D-1}{r}\,\pa_r - \frac{\ell\,(\ell+D-2)}{r^2} \right)\
+\ V(r) \right]\,\Psi_{n_r,\ell}(r)\ =\ E_{n_r,\ell}\,\Psi_{n_r,\ell}(r)\ ,\  \pa_r \equiv \frac{d}{dr}\ .
\label{radial}
\end{equation}
For any radial potential, the energy $E_{n_r,\ell}$ of excited state is always degenerate with respect
to the quantum numbers $\{m_{\ell}\}$. Specifically, for given $n_r$, $\ell$ and $D>1$ we have $\mathcal{N}(D,\ell)$ different wave functions with the same energy. For this reason since now on
we omit the label $\{m_{\ell}\}$ in the energy and the wave function.

The ground state is non-degenerate (and nodeless), it has to zero angular momentum $\ell=0$  thus being $S$-state and zero radial quantum number $n_r=0$. The angular part corresponds to the zero harmonics (which is a constant) with zero eigenvalue in (\ref{sphlap}). From (\ref{radial}) it can be seen that the ground wave function (as well as all $S$ state eigenfunctions) depends only on radial coordinate $r$ that now on we denote by $\Psi(r)$ without labels. It corresponds to the lowest energy eigenfunction of the radial Schr\"odinger operator
\begin{equation}
  \hat{h}_r\ =\ -\frac{\hbar^2}{2M}\left(\pa_r^2\ +\ \frac{D-1}{r}\,\pa_r\right)\ +\ V(r)\ ,
\label{radialop}
\end{equation}
for which the corresponding eigenvalue equation with boundary conditions read
\begin{equation}
    \hat{h}_r\,\Psi(r)\ =\ E\,\Psi(r)\ ,\qquad \int_0^{\infty}\Psi^2\,r^{D-1}dr
    \ <\ \infty\ ,\qquad \Psi(0)\ =\ 1,\quad \Psi(\infty)\ =\ 0\ .
\label{schroedingerR}
\end{equation}
Here $E$ denotes the ground state energy, assuming that $\hbar,m$ are dimensionless, it has dimension $[cm]^{-2}$ as well as the potential. From now on $D$ can be considered as a continuous parameter. Interestingly, some non-trivial analytical properties of the ground state energy as a function of $D$ appear
at both non-physical dimensions $D \leq 0$ and in the physical ones, $D > 0$, see e.g. \cite{DOLGOVPOPOV1979} and \cite{DOREN}.

Taking the ground state wave function $\Psi$ in the exponential representation (\ref{exponentialrep}) and substituting it into the eigenvalue problem (\ref{schroedingerR}) with radial operator (\ref{radialop}) we arrive at the Riccati equation
\begin{equation}
 \hbar\,\pa_ry(r)\ -\ y(r)\,\left(y(r) - \frac{\hbar(D-1)}{r}\right)\ =\ 2 M\, \bigg[E\ -\  V(r)\bigg]\ ,
\label{riccatigeneral}
\end{equation}
where function $y(r)$ is the logarithmic derivative of $\Psi(r)$ or, equivalently, the derivative of the phase $\Phi(r)$,
\begin{equation}
\label{y}
     y(r)\ =\ -\hbar\,\pa_r\left(\log\Psi(r)\right)\ =\ -\ \hbar\,\frac{\pa_r\Psi(r)}{\Psi(r)}\ =\ \pa_r\Phi(r)\ .
\end{equation}
We focus on case when the potential $V(r)$ has the form of finite-degree polynomial in $r$,
\begin{align}
  V(r\,; a_2, a_3,... , a_m)\ & =\ \frac{1}{g^2}\,\sum_{k=2}^{m}a_k\,g^k\,r^k\ ,
\label{potential}
\end{align}
see (\ref{Hamiltonian}). Hence, the Riccati equation reads
\begin{equation}
  \hbar\,\pa_ry(r)\ -\ y(r)\,\left(y(r) - \frac{\hbar(D-1)}{r}\right)\ =2M\left[E\ -\  \frac{1}{g^2}\hat{V}(gr)\right]\ ,
\label{riccatikey}
\end{equation}
c.f. (\ref{riccati1d}).
This is the basic equation in the present work.

\subsection{Riccati-Bloch Equation}

The Riccati equation (\ref{riccatikey}) can be transformed into one without explicit dependence on
$\hbar$ and $M$ by introducing new variable
\begin{equation}
\label{changev}
     v\ =\ \bigg(\frac{2M}{\hbar^2}\bigg)^{\frac{1}{4}} \,r\ ,
\end{equation}
and making the replacements
\begin{equation}
\label{changes}
 y\ =\ (2M \hbar^2)^{\frac{1}{4}}\ \mathcal{Y}\left(v\right)\quad ,\qquad\quad E\ =\ \frac{\hbar}{(2M)^{\frac{1}{2}}}\,\veps\ .
\end{equation}
Then, the equation (\ref{riccatikey}) becomes
\begin{equation}
\label{riccati-bloch}
 \pa_v\mathcal{Y}(v)\ -\ \mathcal{Y}(v)\left(\mathcal{Y}(v)-\frac{D-1}{v}\right)\ =\ \veps\left(\la\right)\ -\ \frac{1}{\la^2}\,\hat{V}\left(\la v\right)\ ,
 \quad \pa_v \equiv \frac{d}{dv}\ ,
\end{equation}
where
\begin{equation}
\label{effective}
 \la\ =\ \left(\frac{\hbar^2}{2M}\right)^{\frac{1}{4}}\, g\ ,
\end{equation}
plays the role of the \textit{effective} coupling constant replacing $g$. It is evident that $\la v = g r$.
The boundary conditions those we impose in equation (\ref{riccati-bloch}) are
\begin{equation}
 \pa_v\mathcal{Y}(v)\Bigr|_{v=0}\ =\ \frac{\veps}{D}\quad , \quad
 \mathcal{Y}(+\infty)\ =\ +\infty\ .
\end{equation}
Evidently, if $\hbar=1$ and $M=1/2$, the equation (\ref{riccati-bloch}) coincides
with (\ref{riccatikey}) and $\la=g$.
We call the equation (\ref{riccati-bloch}) the Riccati-Bloch (RB) Equation.

The RB equation (\ref{riccati-bloch}) allows us to construct the asymptotic expansions
of $\mathcal{Y}(v)$, and ultimately of $y(r)$, at small and large $v$.
For fixed $\la \neq 0$ and small $v$ we have the Taylor series
\begin{equation}
 \mathcal{Y}(v)\ =\
 \frac{\veps }{D}\,v + \frac{\left(\veps^2-a_2D^2\right)}
  {D^2 (D+2)}\,v^3 - \frac{ a_3 \la}{D+3}\,v^4 - \frac{a_4  D^3(D+2) \la^2-2\,\veps\,
  (\veps^2-a_2D^2)}{D^3 (D+2) (D+4)}\,v^5\ +\ \ldots \ ,
\label{yrsmall}
\end{equation}
while for large $v$,
\[
  \mathcal{Y}(v)\ =\ a_m^{1/2} \la^{(m-2)/2} v^{m/2}\ +\
\]
\begin{equation}
 \frac{ a_{m-1} \la^{(m-4)/2}}{2 a_m^{1/2}}v^{(m-2)/2}\ +\
   \frac{ (4a_ma_{m-2}\ -\ a_{m-1}^2) \la^{(m-6)/2}}{8 a_m^{3/2}}v^{(m-4)/2}\ +\ \ldots \ .
\label{asymptoticgeneral}
\end{equation}
It is easy to check that for $\la=0$, hence, for the harmonic oscillator case $V=a_2\,r^2$, the equation (\ref{riccati-bloch}) has the exact solution
\begin{equation}
\label{unperturbed0}
   \mathcal{Y}(v) =\ \frac{\veps}{D}\,v\ ,\qquad \veps\ =\ a_2^{1/2}\,D\ .
\end{equation}
Three remarks in a row: $(i)$ the  expansion (\ref{yrsmall}) at $\la=0$ is terminated and consists  of the first term alone, leading to  $\mathcal{Y}(v)=(\veps/D)\,v$, in agreement with (\ref{unperturbed0}), $(ii)$ in general, any coefficient in  (\ref{yrsmall}) depends on energy explicitly,
$(iii)$ the fourth and subsequent (unwritten) terms in (\ref{asymptoticgeneral}) can acquire explicit dependence on  $\veps$. It is worth mentioning that for fixed $M$, large $v$ behavior can be obtained in two situations: large $r$ and fixed $\hbar $, or at fixed $r$ and small $\hbar$, see (\ref{changev}). Hence, the expansion (\ref{asymptoticgeneral}) describes the semiclassical limit $\hbar \rar 0$.

Integrating (\ref{yrsmall}) and  (\ref{asymptoticgeneral}) in $v$,  both expansions are converted into expansions for the phase $\Phi$, namely,
\begin{equation}
 \frac{1}{\hbar}\Phi\ =\ \frac{\veps }{2D}v^2+\frac{\left(\veps^2-a_2 D^2\right) }{4D^2 (D+2)}v^4-\frac{ a_3 \lambda}{5(D+3)}v^5-\frac{a_4  D^3(D+2) \lambda^2-2\veps (\veps^2-a_2D^2)}{6D^3 (D+2) (D+4)}v^6\ +\ \ldots\ ,
\label{phirsmall}
\end{equation}
and
\begin{equation}
 \frac{1}{\hbar}\Phi\ =\ \frac{2a_m^{1/2}}{m+2}  \lambda^{(m-2)/2} v^{(m+2)/2}+\frac{ a_{m-1}  \lambda^{(m-4)/2}}{m\, a_m^{1/2}}v^{m/2}\ +\ \frac{ (4a_ma_{m-2}-a_{m-1}^2)  \lambda^{(m-6)/2}}{4(m-2) a_m^{3/2}}v^{(m-2)/2}\ +\ \ldots\ ,
\label{phirlarge}
\end{equation}
for small and large $v$, respectively. Later these expansions will be used to make interpolation between large and small $v$ to construct a (uniform) approximation for the phase.

\subsection{Generalized Bloch Equation}

There exists an alternative way to transform the original Riccati equation (\ref{riccatikey}) into equation without the explicit dependence on $\hbar$ and $M$. This is achieved by using a change of variable
\begin{equation}
\label{changeu}
    u\ =\ g\,r\ ,
\end{equation}
and introducing a new unknown function
\begin{equation}
  \mathcal{Z}\left(u\right)\ =\ \frac{g}{(2M)^{1/2}}\, y\ ,
\end{equation}
where $\mathcal{Z}(u)$ satisfies the non-linear differential equation
\begin{equation}
 \la^2\,\pa_u\mathcal{Z}(u)\ -\ \mathcal{Z}(u)\left(\mathcal{Z}(u)-\frac{\la^2(D-1)}{u}\right)\ =\ \la^2\,\veps(\la)\ -\ \hat{V}(u)\quad ,\quad \pa_u\equiv\frac{d}{du}\ ,
\label{Bloch}
\end{equation}
c.f.(\ref{riccati-bloch}), with boundary conditions
\begin{equation}
 \pa_u\mathcal{Z}(u)\Bigr|_{u=0}\ =\ \frac{\veps}{D}\quad ,\quad
 \mathcal{Z}(+\infty)\ =\ +\infty\ .
 \label{boundary}
\end{equation}
Here $\veps$ and $\la$ are the same as in (\ref{changes}) and (\ref{effective}). They play the role of energy and effective coupling constant in both $v$-space (\ref{changev}) and $u$-space (\ref{changeu}) dynamics.

At $D=1$ the equation (\ref{Bloch}) was called in \cite{ESCOBARI,ESCOBARII,Shuryak} the \textit{(one-dimensional) GB Equation}. Here, the equation (\ref{Bloch}) is a natural extension of one-dimensional GB equation to $D$-dimensional case, for a $D$-dimensional radial potential. For this reason we continue to call it the {\it Generalized (radial) Bloch} Equation.  Naturally, (\ref{Bloch}) can also be used to construct the asymptotic expansions similar to (\ref{yrsmall}) and (\ref{asymptoticgeneral}) for small $u$,
\begin{equation}
\mathcal{Z}(u)\ =\ \frac{\veps }{D}\,u + \frac{\left(\veps^2-a_2D^2\right)}
  {D^2 (D+2)\lambda^2}\,u^3 - \frac{ a_3 }{(D+3)\lambda^2}\,u^4 - \frac{a_4  D^3(D+2) \la^2-2\,\veps\,
  (\veps^2-a_2D^2)}{D^3 (D+2) (D+4)\lambda^4}\,u^5\ +\ \ldots \ ,
\end{equation}
and large $u$,
\begin{equation}
 \mathcal{Z}(u)\ =\ a_m^{1/2}\, u^{m/2}\ +\
 \frac{ a_{m-1}}{2 a_m^{1/2}}\,u^{(m-2)/2}\ +\
 \frac{ (4a_ma_{m-2}\ -\ a_{m-1}^2)}{8 a_m^{3/2}}\,u^{(m-4)/2}\ +\ \ldots \ ,
\end{equation}
respectively.

\section{The Weak Coupling Regime}

\label{Weak Coupling}

In this Section we describe how to solve both the RB equation and the GB equation using PT in powers of the effective coupling constant $\la$. In particular, for the RB equation the perturbative solution can be obtained applying the so-called Non-Linearization Procedure \cite{TURBINER:1984}. Due to the importance of this procedure, we begin Section by giving a brief description of this method taken as example the $D$-dimensional radial potential.

\subsection{The Non-Linearization Procedure}
\label{PT}

Let us take the RB equation (\ref{riccati-bloch})
\begin{equation}
\label{riccatigeneral-1}
   \pa_v\mathcal{Y}(v)\ -\ \mathcal{Y}(v)\left(\mathcal{Y}(v)-\frac{D-1}{v}\right)\ =\ \veps\left(\lambda\right)\ - V(v; \la)\ ,
\end{equation}
with potential $V(v; \la)$ which admits a Taylor expansion
\begin{equation}
V(v;\la)\ =\ \sum_{n=0}^{\infty}V_n(v)\,\la^n\ .
\label{potentialp}
\end{equation}
Here $\la$ is formal parameter and the coefficient functions $V_n(v)$, $n=0,1,...,$  are real functions in $v$. Let us now assume that the unperturbed equation at $\la=0$,
\begin{equation}
   \pa_v\mathcal{Y}_0(v)\ -\ \mathcal{Y}_0(v)\,\left(\mathcal{Y}_0(v)-\frac{D-1}{v}\right)\ =\ \veps_0\ -\ V_0(v)\ ,
\label{unperturbed}
\end{equation}
can be solved explicitly. It can always be achieved via the \textit{inverse problem}: we take some function $\mathcal{Y}_0(v)$ and then calculate r.h.s.: the potential $V_0(v)$ and  $\veps_0$. Evidently, once we know $\mathcal{Y}_0(v)$, the wave function can be found,
\begin{equation}
    \Psi_0(v)\ =\ \exp\left(- \int^{v} \mathcal{Y}_0(s)\,ds \right)\ .
\label{PSI0}
\end{equation}
It leads to a constraint on the choice of $\mathcal{Y}_0(s)$: the function $\Psi_0(v)$ should be normalizable.
Now, we can develop PT in powers of $\la$,
\begin{equation}
   \veps(\la)\ =\ \sum_{n=0}^{\infty} \veps_n\,\la^n
\label{seriespE}
\end{equation}
and
\begin{equation}
   \mathcal{Y}(v)\ =\ \sum_{n=0}^{\infty}\mathcal{Y}_n(v)\,\la^n\ .
\label{seriespy}
\end{equation}
Substituting (\ref{seriespE}) and (\ref{seriespy}) into the RB equation (\ref{riccatigeneral-1}) it is easy to see that the $n$th correction $\mathcal{Y}_n(v)$ satisfies the first order linear differential equation
\begin{equation}
 \pa_v \left( \,v^{D-1}\Psi_0^2\ \mathcal{Y}_n(v)\right)\ =\
 \left( \veps_n-Q_n(v)\right)v^{D-1}\Psi_0^2 \ \ ,
\label{RiccatiPert}
\end{equation}
where
\begin{equation}
Q_1(v)\ =\ V_1(v)\ ,\quad\quad\quad Q_n(v)\ =\ V_n(v)\ -\ \sum_{k=1}^{n-1}\mathcal{Y}_k(v)\,\mathcal{Y}_{n-k}(v)\ ,\quad n=2,3, \ldots\ .
\label{qn}
\end{equation}
Evidently, the solution for $\mathcal{Y}_n(v)$ is given by
\begin{equation}
 \mathcal{Y}_n(v)\ =\ \frac{1}{v^{D-1}\,\Psi_0^{2}}\left( \int_ 0^{v}(\veps_n-Q_n)\,\Psi_0^2\,s^{D-1}\,ds\right)\ .
\label{ycorrection-gen}
\end{equation}
Since we are interested in finding bound states we have to impose the condition of the absence of current of particles for both $v \rightarrow 0$ and  $v\rightarrow \infty$ as  boundary condition \cite{TURBINER:1984},
\begin{equation}
\mathcal{Y}_n(v)\,v^{D-1}\,\Psi_0^2\Bigr|_{v=\{0,\infty\}}\ \rar \ 0 \ .
\label{Boundary}
\end{equation}
It can be easily checked that if $v$ tends to zero the condition (\ref{Boundary}) is satisfied
automatically while at $v=\infty$ the correction $\veps_n$ should be chosen accordingly
\begin{equation}
  \mathcal{Y}_n(v)\ =\ \frac{1}{v^{D-1}\,\Psi_0^{2}}
  \int_0^{v}(\veps_n-Q_n)\,\Psi_0^2\,s^{D-1}\,ds \ ,
\label{ycorrection}
\end{equation}
where
\begin{equation}
   \veps_n\ =\ \frac{\int_0 ^\infty Q_n\,\Psi_0^2\,v^{D-1}\, dv}
     {\int_0 ^\infty \Psi_0^2\,v^{D-1}\,dv}  \ .
\label{ecorrection}
\end{equation}
to satisfy (\ref{Boundary}).
For $D=1$, this perturbative approach was called the {\it Non-Linearization Procedure} \cite{TURBINER:1984}. Here we have given the straightforward extension of it to radial potentials in arbitrary $D$. We will continue to call it the {\it Non-Linearization Procedure}.
In contrast with the Rayleigh-Schr\"odinger PT, the knowledge of entire spectrum of the unperturbed problem is not required to find constructively perturbative corrections in (\ref{seriespE}) and (\ref{seriespy}). It is sufficient to know the unperturbed ground state wave function of the unperturbed problem to which we are looking for corrections.
This approach gives the closed analytic expression for both corrections $\veps_n$ and $ \mathcal{Y}_n(v)$ in form of nested integrals.  Therefore, this procedure is the efficient method to calculate several orders in PT, see e.g. \cite{TURBINER:1984} and \cite{delValle}.
Interestingly, in this framework the convergence of the perturbation series (\ref{seriespE})
and (\ref{seriespy}) is guaranteed if the first correction $\mathcal{Y}_1(v)$ is bounded
\begin{equation}
| \mathcal{Y}_1(v)|\ \leq\ \textit{Const}
\end{equation}
for any $D$, for discussion see \cite{TURBINER:1984}.

We have presented a brief review of the Non-Linearization Procedure applied to the ground state. This approach can be modified to study the excited states by admitting a number of simple poles with residues equal to one in $\mathcal{Y}_n(v)$. The pole positions are found in PT in $\la$. Explicit formulas can be found in \cite{TURBINER:1984} for one-dimensional case. The generalization for the $D$-dimensional radial potential case is straightforward.

\subsection{The Weak Coupling Expansion from the Riccati-Bloch Equation}

Now, let us consider the $D$-dimensional radial anharmonic oscillator potential (\ref{potential}) in weal coupling regime assuming that the effective  coupling constant
$\la = \left(\frac{\hbar^2}{2M}\right)^{\frac{1}{4}}\,g$ is \textit{small}.
In order to apply the Non-Linearization Procedure of Subsection \ref{PT}
to construct the weak coupling expansion we choose in (\ref{potentialp})
\begin{eqnarray}
\label{selection}
  \la & \ =\ & \left(\frac{\hbar^2}{2M}\right)^{\frac{1}{4}}\, g\ , \non \\
  V_k & \ =\ & a_{k+2}\,v^{k+2}\quad \mbox{for} \quad k=0,\,1,\ldots, m-2\ ,\ \mbox{and} \non \\
  V_k &   =  & 0\ \quad\text{for}\quad k>m-2\  ,
\end{eqnarray}
hence, our potential is a finite degree polynomial in both $\la$ and $v$.
The corresponding unperturbed equation (\ref{unperturbed}) describes the $D-$dimensional spherical harmonic oscillator: an exactly solvable problem in any $D > 0$, see e.g. \cite{LYNCH1990127}. The unperturbed ground state wave function is given by
\begin{equation}
\Psi_0\ =\ e^{-\sqrt{\frac{M}{2\hbar^2}}\ a_2^{1/2}r^2}\ =\ e^{-\frac{1}{2}a_2^{1/2}v^2}\ .
\end{equation}

In general, the explicit calculation of the perturbative corrections $\mathcal{Y}_n(v)$ and
$\veps_n$ can be carried out analytically by using (\ref{ycorrection}),
final expressions involve explicitly the incomplete gamma function.
Let us find the asymptotic expansions of $n$th correction $\mathcal{Y}_n(v)$ in two limits:
$v \rar 0$ and $v\rar \infty$. For small $v$ the correction $\mathcal{Y}_n(v)$ is given by
the Taylor expansion
\begin{equation}
  \mathcal{Y}_n(v)\ =\ v\sum_{k=0}^{\infty}b_k^{(n)}\,v^k\ ,
\label{small}
\end{equation}
where $b_k^{(n)}$ are some coefficients. It can be easily shown that the first coefficient
$b_0^{(n)}$ is related to the energy correction $\veps_n$,
\begin{equation}
\label{b0}
b_0^{(n)}\ =\ \frac{\veps_n}{D}\ ,
\end{equation}
while the next coefficient always vanishes,
\begin{equation}
    b^{(n)}_1\ =\ 0\ ,
\end{equation}
due to the absence of the linear term in the potential (\ref{potential}).
As for large $v$ at $n > 0$ we have a Laurent series expansion,
\begin{equation}
    \mathcal{Y}_n(v)\ =\ v\,\sum_{k=0}^{\infty}c^{(n)}_{k}\,v^{n-k} \ ,
\label{large}
\end{equation}
where $c_k^{(n)}$ are some coefficients. Interestingly, the leading coefficient $c_0^{(n)}$ does not depend on $D$ while the next-to-leading one always vanishes,
\begin{equation}
    c^{(n)}_1\ =\ 0\ .
\label{c_1}
\end{equation}

At $D = 0$ all corrections $\veps_n$ vanish,
\[
   \veps_n\ =\ 0 \ .
\]
It can be demonstrated by using the expression (\ref{ecorrection}) for $\veps_n$, and the asymptotic expansions (\ref{small}) and (\ref{large}). When $D \rar 0$ the numerator of (\ref{ecorrection}) is bounded,
\begin{equation}
   \int_0^\infty Q_n\,\Psi_0^2\,v^{D-1}\,dv\ <\ \infty\ ,
\end{equation}
while  the denominator has the asymptotic series expansion
\begin{equation}
    \int_0^\infty \Psi_0^2\,v^{D-1}\,dv\ =\ \frac{a_2^{-\frac{D}{4}}}{D}\ +\ O(D^0)\ .
\end{equation}
Consequently, when $D \rar 0$ the correction $\veps_n$ vanishes linearly, the coefficient $b_0^{(n)}$ remains finite, see (\ref{b0}). If $\veps_n$ vanishes for all $n$, then their formal sum (\ref{seriespE}) results in $\veps=0$ and, ultimately, $E=0$.
In general, for $D \neq 0$ it can be shown that expansion for $\veps$ is asymptotic: $\veps_n$ grows factorially as $n \rar \infty$, see e.g. \cite{BENDER,BENDERII,BANKS,SILVERSTONE}. Therefore, series (\ref{seriespE}) and (\ref{seriespy}) are divergent in $\la$. For some particular cases of (\ref{Hamiltonian}), it has been proved that perturbation series for the energy can be summed to the exact one by using different techniques:  calculating the Borel sum \cite{GRAFFI}, taking Pad\' e approximants \cite{LOEFFEL} and via renormalization \cite{KILLINGBECK}.

An interesting situation occurs when all odd monomial terms in $r$ in potential (\ref{potential}) are absent, i.e. the potential is (formally) an even function, $V(r)=V(-r)$. In this case, all odd corrections  $\mathcal{Y}_{2n+1}(v)$ and $\veps_{2n+1}$ vanish. The even correction $\mathcal{Y}_{2n}(v)$  has the form of a polynomial of finite  degree
\begin{equation}
  \mathcal{Y}_{2n}(v)\ =\ v \sum_{k=0}^{n} c^{(2n)}_{2k}v^{2(n-k)}\ ,\quad\ \mathcal{Y}_{2n}(-v)\ =\ -\mathcal{Y}_{2n}(v)\ ,
\label{polinomialyn}
\end{equation}
where
\begin{equation}
  c_{2n}^{(2n)}\ =\ \frac{\veps_{2n}}{D}\ .
\end{equation}
This implies that $\veps_{2n}$  and $\mathcal{Y}_{2n}(v)$ can be calculated by linear algebra means. In turn, the energy correction $\veps_{2n}$ is a finite-degree polynomial in $D$,
\begin{equation}
   \veps_{2n}=\sum_{k=1}^{n+1}d_k^{(2n)}D^k\ ,
\label{polinomialEn}
\end{equation}
where $d_k^{(2n)}$ are real coefficients. However, it is enough to have a single odd monomial term $a_{2q+1} v^{2q+1}$ in the potential (\ref{potential}) (naturally, $q$ is integer) to break the features (\ref{polinomialyn}) - (\ref{polinomialEn}). Hence, all coefficients in front of the singular terms in the expansion (\ref{large}), as well as all higher order terms in the expansion (\ref{small}) at $k > n$, are  proportional to $a_{2q+1}$.

\subsection{The Weak Coupling Expansion from the Generalized Bloch Equation}

Taking in the GB equation (\ref{Bloch}) the effective coupling constant
$\la = \left(\frac{\hbar^2}{2M}\right)^{\frac{1}{4}}\,g$ as a formal parameter we can develop PT in powers of $\la$. The expansion of the eigenvalue $\veps$ in powers of $\la$ remains the same as for the RB equation, see (\ref{seriespE}),
\begin{equation*}
   \veps(\la)\ =\ \sum_{n=0}^{\infty}\veps_n\,\la^n\ ,
\end{equation*}
while $\mathcal{Z}(u)$ is of the form of the Taylor series
\begin{equation}
 \mathcal{Z}(u)\ =\ \sum_{n=0}^{\infty} \mathcal{Z}_n(u)\,\la^n\ .
\label{seriesZ}
\end{equation}
We assume that the perturbative energy corrections $\veps_n$ are already known: they can be found using the Non-Linearization Procedure, the standard Rayleigh-Schr\"odinger PT or any other suitable method. It is immediate to see that correction $\mathcal{Z}_n(u)$ is calculated by algebraic means and depends on corrections of smaller order,
\begin{align}
 \mathcal{Z}_0(u) & \ =\ \pm\,\sqrt{\hat{V}(u)}\  ,  \non \\
 \mathcal{Z}_1(u)& \ =\  0\ ,\non \\
 \mathcal{Z}_2(u) & \ =\  \frac{u\,\pa_u \mathcal{Z}_{0}(u)\ +\ (D-1)\,\mathcal{Z}_{0}(u)\ -\ u\,\veps_0}{2\,u\,\mathcal{Z}_{0}(u)}\  , \non\\
\vdots\non\\
 \mathcal{Z}_n(u) & \ =\  \frac{u\,\pa_u \mathcal{Z}_{n-2}(u)\ +\ (D-1)\,\mathcal{Z}_{n-2}(u)\ -\ u^2\sum_{i=2}^{n-2}\mathcal{Z}_{i}(u)\mathcal{Z}_{n-i}(u)\ -\ u\,\veps_{n-2}}{2\,u\,\mathcal{Z}_{0}(u)} \quad
\text{for} \quad n>2 \ .
\label{BlochSol}
\end{align}
Note that the boundary condition $\mathcal{Z}(\infty)\,=\,+\infty$ implies the positive sign in the expression for $\mathcal{Z}_0(u)$ should be chosen.

Making the analysis of $\mathcal{Z}_n(u)$ correction (\ref{BlochSol}) one can see that the boundary condition  at $u=0$, see (\ref{boundary}), can {\it not} be fulfilled in the case of arbitrary odd anharmonic radial potential (\ref{potential}) once the expansion (\ref{seriesZ}) for $\mathcal{Z}(u)$ is used. In general, any $\mathcal{Z}_n(u)$ (and its derivative) at $n>1$ diverges at small $u$. For instance, for the cubic potential where $a_3 \neq 0$ the $n$th correction $\mathcal{Z}_n(u)$ with $n>1$ behaves like
\begin{equation}
 \mathcal{Z}_n(u) \sim u^{-n+2}
\end{equation}
when $u \rar 0$. In turn, for the quintic potential with $a_3=0$ but $a_5 > 0$ the $n$th correction at $n>3$ behaves like
\begin{equation}
  \mathcal{Z}_n(u)\sim u^{-n+4}
\end{equation}
when $u$ tends to zero. For the polynomial in $u$ potential of degree $(2k+1)$, if all other odd terms are absent $a_3=a_5=\ldots=a_{2k-1}=0$ but $a_{2k+1} > 0$,
the function $\mathcal{Z}_n(u)$ behaves like
\begin{equation}
 \mathcal{Z}_n(u)\sim u^{-n+2k}
\end{equation}
for small $u$ as long as $n>2k-1$. It is the clear indication that the radius of convergence of the expansion of $\mathcal{Z}(u)$ (see (\ref{seriesZ})) in $1/u$ is finite.
However, if the anharmonic potential $V(r)$ is even, $V(r)=V(-r)$ the boundary condition at $u=0$ can be satisfied. It might be an indication that the radius of convergence in $1/u$ is infinite. It allows us to determine the correction $\veps_n$ by imposing the boundary condition
\begin{equation}
\label{zn0}
\pa_u\mathcal{Z}_n(u)\Bigr|_{u=0}\ =\ \frac{\veps_n}{D}\ .
\end{equation}

\begin{figure}[h]
  \includegraphics[width=0.7\columnwidth]{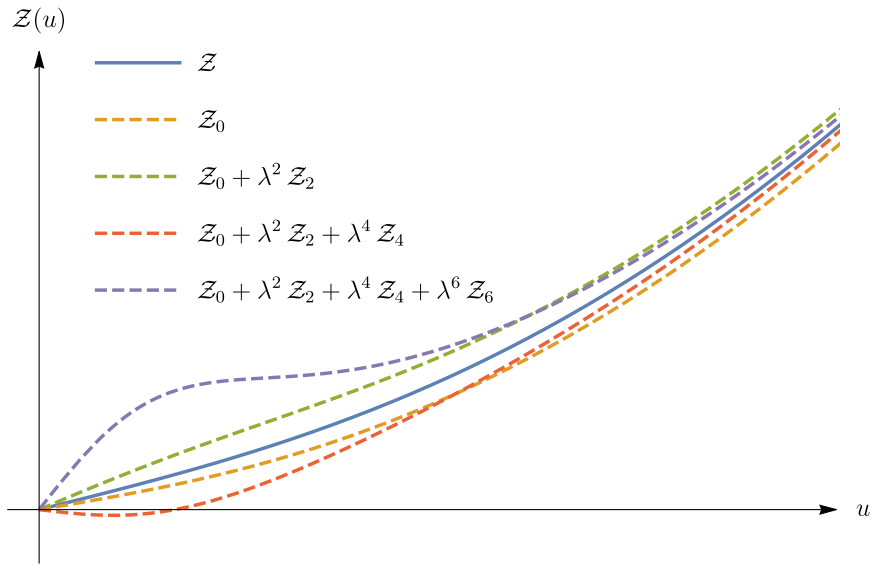}
  \caption{Partial sums of the expansion for $\mathcal{Z}(u)$ as functions of $u$ for a general  anharmonic potential with monomials of even degrees. The function $\mathcal{Z}(u)$ (solid blue line), represents the exact solution. Near $u=0$, the deviation of the partial sums from the  exact solution is a consequence of the divergent nature of series (\ref{seriespy}).}
   \label{fig:even}
\end{figure}

\begin{figure}[h]
   \includegraphics[width=0.7\columnwidth]{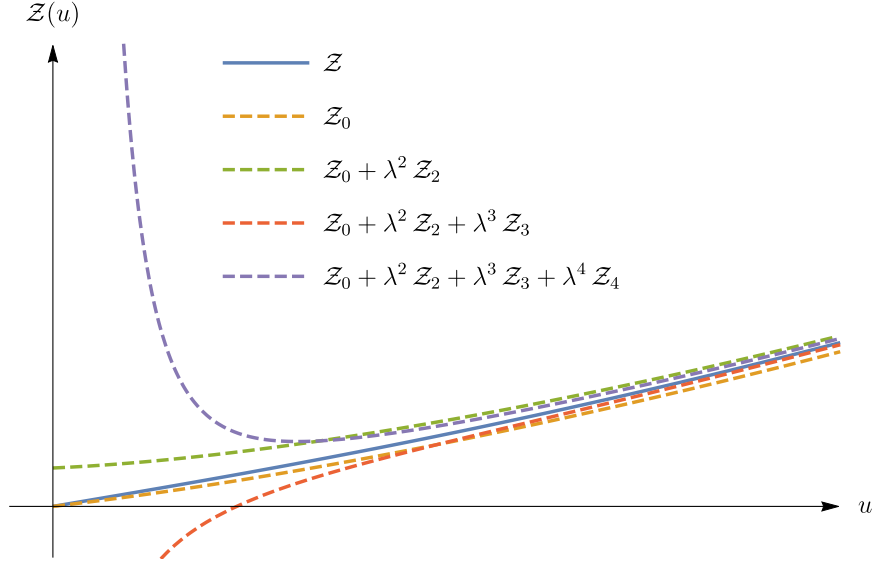}
   \caption{Partial sums of the expansion for $\mathcal{Z}(u)$ as functions of $u$ for a generic  anharmonic potential which includes odd monomials. The function $\mathcal{Z}(u)$  (solid blue line)  represents the exact solution.  The divergence in vicinity of $u=0$ of the partial sums is the result of the impossibility of satisfying the boundary condition at $u=0$.}
    \label{fig:odd}
\end{figure}

At this point,  it is worth remarking that the expansion in powers of $\la$ (\ref{seriesZ}) as well as (\ref{seriespy}) is divergent for any anharmonic oscillator due to the so-called Dyson instability, see \cite{Dyson} and as for the discussion \cite{TURBINER:1984}. This fact can be seen in the partial sums of  (\ref{seriesZ}), see Figs. \ref{fig:even} and \ref{fig:odd}.

Let us emphasize that the presence of a single odd degree monomial term in $u$ in the potential $V(u)$ breaks the feature (\ref{zn0}): the expansion for $\mathcal{Z}(u)$ in powers of $\la$ only satisfies the single boundary condition at $u=\infty$. To explain why this happens, it is  necessary to make a connection of the expansions (\ref{seriespy}) and (\ref{seriesZ}).

\subsection{Connection between $\mathcal{Y}\ \mbox{and}\ \mathcal{Z}$ expansions}

\begin{figure}[h]
   \includegraphics[width=1\textwidth]{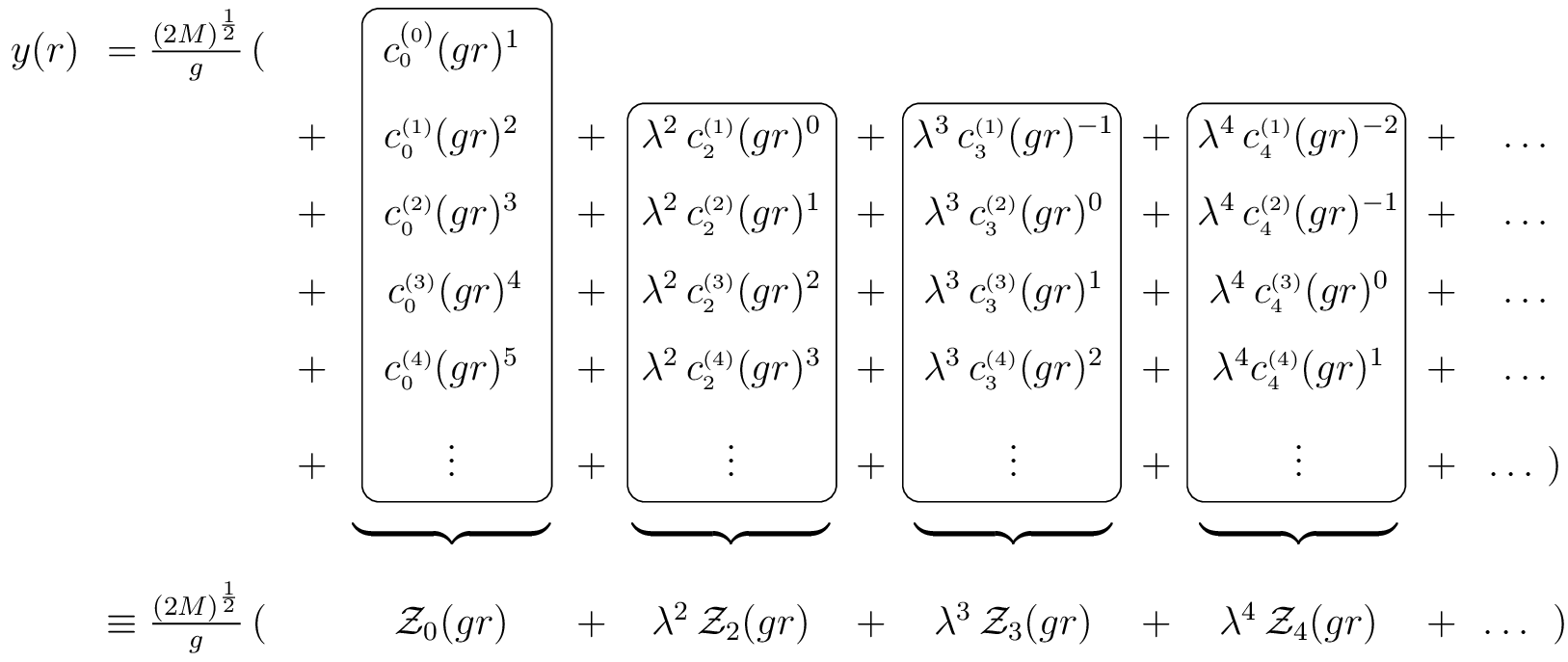}
   \caption{Representation of the generating functions $\mathcal{Z}_n$ constructed
   from the perturbative series in powers of  $\la$ for the function $y(r)$.}
    \label{fig:PTdiagram}
\end{figure}

Thus far we have constructed two different representations for the logarithmic derivative $y(r)$, see (\ref{y}). From one side we have
\begin{equation}
\label{1strep}
   y(r)\ =\ (2M \hbar^2)^{\frac{1}{4}}\,\mathcal{Y}(v(r))\ ,
   \qquad
   \mathcal{Y}(v)\ =\ \sum_{n=0}^{\infty}\mathcal{Y}_n(v)\, \la^n\ ,
   \qquad
   v(r)\ =\ \left(\frac{2M}{\hbar^2}\right)^{\frac{1}{4}}\, r\ ,
\end{equation}
from the another side
\begin{equation}
\label{2ndoption}
    y(r)\ =\ \left(\frac{2M}{g^2}\right)^{\frac{1}{2}}\,\mathcal{Z}(u(r))\ ,
    \qquad
    \mathcal{Z}(u;\la)\ =\ \sum_{n=0}^{\infty} \mathcal{Z}_n(u)\,\la^n\ ,
    \qquad
     u(r)\ =\ g\,r\ ,
\end{equation}
where the coupling constant $\la$ is defined in (\ref{effective}). The connection between them can be established if we use explicitly the expansion of $\mathcal{Y}_n(v)$ at large $v$, see (\ref{large}). In this case $y(r)$ is given by
\begin{equation}
\label{2ndstep}
     y \ =\ (2M\,\hbar^2)^{\frac{1}{4}}\,v\sum_{n=0}^{\infty}\la^n \sum_{k=0}^{\infty}c^{(n)}_{k}\,v^{n-k}\ ,
\end{equation}
or, equivalently
\begin{eqnarray}
 y \ &=&\
      \left(\frac{2M}{g^2}\right)^{\frac{1}{2}}\,\sum_{n=0}^{\infty}\la^n (g r) \sum_{k=1}^{\infty}c_n^{(k)}(gr)^{k-n}\ .
\label{alternative}
\end{eqnarray}
Comparing the expansions (\ref{2ndstep}) and (\ref{alternative}) we can conclude that
\begin{equation}
\label{Z(u)}
\mathcal{Z}_n(u)\ =\ u\,\sum_{k=1}^{\infty}c_n^{(k)}u^{k-n}
\end{equation}
due to the uniqueness of the Taylor series. Therefore, $\mathcal{Z}_n(u)$ is a generating function (!) for the coefficients $c_k^{(n)}$, $k=0,1,...$\,, see for graphical illustration Fig.\ref{fig:PTdiagram}.
Equation (\ref{Z(u)}) displays the reason why we cannot satisfy, in general, the boundary condition at $u=0$: $\mathcal{Z}_n(u)$ is constructed from the expansion (\ref{large}) which is valid in the limit $r\rar\infty$ ($u\rar\infty$ for fixed $g$). In fact, some interesting properties of the corrections $\mathcal{Z}_n(u)$ occur when considering $r\rar\infty$. For example, the first term in  (\ref{2ndoption}) has the expansion
\begin{eqnarray}
 \frac{\mathcal{Z}_0(gr)}{g}\,&\ =&\ a_m^{1/2}\, g^{(m-2)/2} r^{m/2}\ +\ \frac{ a_{m-1}\, g^{(m-4)/2}}{2 a_m^{1/2}}r^{(m-2)/2} \ \non \\
 && +\ \frac{ (4a_ma_{m-2}\ -\ a_{m-1}^2)\, g^{(m-6)/2}}{8 a_m^{3/2}}r^{(m-4)/2}\ +\ \ldots \ .
\end{eqnarray}

This expansion reproduces exactly the dominant asymptotic behavior of the function $y(r)$ up to the term $O\left(r^{(m-4)/2}\right)$, c.f. (\ref{asymptoticgeneral}). Note that the next generating function  $\mathcal{Z}_1(u)=0$ in agreement with (\ref{c_1}). It is evident that the next generating functions
in (\ref{asymptoticgeneral}): $\mathcal{Z}_2(u)$, $\mathcal{Z}_3(u)$, ..., also
contribute to the expansion of $y(r)$. It will be discussed in detail for the cubic anharmonic oscillator, see below.

Let us conclude this Section by clarifying the perturbative approach for the GB equation and its connection with the semiclassical WKB approximation. By integrating in $r$,  the expansion (\ref{2ndoption}) can be converted into an expansion for the phase
\begin{equation}
\label{semiclassical}
 \Phi(r;\la)\ =\ \sum_{n=0}^{\infty}\la^n G_n(r)\quad ,\qquad G_1(r)\ =\ 0\ ,
\end{equation}
where $G_n(r)$ is equal to
\begin{equation}
  G_n(r)\ =\ \left(\frac{2M}{g^2} \right)^{1/2}\,\int^{r}\mathcal{Z}_n(gr)\,dr\ .
\label{semiclassicalcorrection}
\end{equation}
Note that keeping $g$ fixed, (\ref{semiclassical}) can be regarded as a semiclassical expansion of the phase in powers of $\hbar^{\frac{1}{2}}$, see (\ref{effective}).  The leading order term is
\begin{equation}
\label{G0}
      G_0(r)\ =\ \int^{r}\sqrt{2M \,V(r)}\,dr\ ,
\end{equation}
for $V \geq 0$ it is the classical action at $E=0$ and it corresponds to the zero order
of the standard one-dimensional WKB method (developed in the classically forbidden region).
This is the only term in expansion (\ref{semiclassical}) which contains no dependence on $D$.
The second  order term in (\ref{semiclassical}) is
\begin{equation}
\label{second}
 \la^2 G_2(r)\ =\ \hbar\left(\frac{D-1}{2}\log[ r] +\ \frac{1}{4}\,\log\left[
  2M\,V(r)\right]\ -\ \frac{\veps_0}{2}\int^{r}\frac{dr}{\sqrt{V(r)}}\right)\ .
\end{equation}
Except for the integral - the third term - it looks like the first order correction to WKB at $E=0$ and $D=1$. Thus, it defines the determinant \cite{ESCOBARII}. The appearance of the extra term in the form of integral can be explained as follows. Let us take the zero order term in the standard WKB method (in the forbidden region) and expand it in powers of $\hbar^{1/2}$ using (\ref{changes}) and (\ref{effective}), thus
\begin{equation}
 \int^{r}\sqrt{2M\left(V(r)-E\right)}\,dr\ =\ \int^{r}\sqrt{2M\,V(r)}\,dr\ -\  \frac{\hbar\,\veps_0}{2}\int^{r}\frac{dr}{\sqrt{V(r)}}\ +\ O(\hbar^{3/2})\ .
\end{equation}
Note that the first term is nothing but $G_0(r)$ while the second term is exactly the integral in  (\ref{second}). Consequently,  one can see that (\ref{semiclassical}) is the standard semiclassical
WKB expansion in the classically-forbidden region, re-expanded in powers of $\hbar^{1/2}$.
Needless to say, higher order generating functions $G_3(r), G_4(r), \ldots$ are related with
higher order corrections in the WKB expansion in a similar way. They define two-, three-, etc loop contributions in the path integral formalism in Euclidian time for the density matrix at coinciding
points.

Note that this is alternative consideration to the one-dimensional case presented in \cite{ESCOBARI,ESCOBARII,Shuryak}. In these papers it was described how to obtain expansion (\ref{semiclassical}) using a path integral formulation (in Euclidean time) considering a special classical solution at zero energy $E=0$ called {\it flucton}. In quite straightforward way this consideration can be extended to the $D$-dimensional radial case, where the flucton path appears in radial direction. Hence, the perturbative approach developed for the GB Equation can be called the {\it true} semiclassical approximation: both $\veps(\la)$, and $\mathcal{Z}(u;\la)$ and  $\Phi(r;\la)$ are expansions in powers $\hbar^{1/2}$.

\section{The Strong Coupling Expansion}

\label{Strong Coupling}

In the previous Section \ref{Weak Coupling} assuming an \textit{small} effective constant $\la=\left(\frac{\hbar^2}{2M}\right)^{\frac{1}{4}}g$ (\ref{effective}) we studied the ground state energy $E$ and the phase of the wave function in the form of a Taylor expansion at $\la=0$. As a result the harmonic oscillator was taken as unperturbed problem.  We  now consider the opposite limit of large $\la$ taking as the unperturbed potential the leading monomial term $a_m g^{m-2} r^m $ in (\ref{potential}). This situation can be achieved if we consider the expansion in fractional powers $1/\la$ - the strong coupling expansion. It implies that all other terms in the potential (\ref{potential}), except for the leading one are taken as perturbation.

\subsection{The Strong Coupling Expansion from the Riccati Equation}
\label{strong}

In order to develop the strong coupling expansion in the framework of the Non-Linearization Procedure, let us first take the radial Schr\"odinger equation  (\ref{schroedingerR}) with the polynomial potential (\ref{potential}) and perform the Symanzik scaling transformation
\begin{equation}
\label{r_to_r.g^m}
         r\ \rar \ r\,g^{-\frac{m-2}{m+2}}\ ,
\end{equation}
with idea to remove the coupling constant in front of the leading term in potential. As the result
we arrive at the radial Schr\"odinger equation,
\begin{equation}
\label{schroedingerST}
  \left[-\frac{\hbar^2}{2M}\left(\pa_r^2\ +\ \frac{D-1}{r}\,\pa_r\right)\ +\ W(r)\right]\,\Psi\ =\ \tilde{E}\,\Psi\quad , \quad
  \tilde{E}\ =\ g^{-2\frac{m-2}{m+2}}\,E\ ,
\end{equation}
with a potential $W(r)$ given by
\begin{equation}
  W(r)\ =\ \frac{1}{\gamma^m}\sum_{k=2}^{m} a_k\,(\gamma\,r)^k\ =\ a_m r^m + \frac{a_{m-1}}{\gamma}\, r^{m-1} + \ldots + \frac{a_2}{\gamma^{m-2}}\,r^2 \equiv \
  \frac{1}{\gamma ^m}\,\hat{W}(\gamma\,r)\ ,
\label{potentialST}
\end{equation}
and rescaled energy $\tilde{E}$, where
\begin{equation}
\label{1 over g}
  \gamma\ =\ g^{\frac{4}{m+2}}\ .
\end{equation}
Note that now the parameter $\gamma^{-1}$ plays the role of the coupling constant for $W(r)$.
By using the exponential representation (\ref{exponentialrep}), the equation (\ref{schroedingerST})
can be transformed into the Riccati equation (\ref{riccatigeneral}) with potential $W(r)$ and energy $\tilde{E}$ , explicitly,
\begin{equation}
  \hbar\,\pa_ry(r)\ -\ y(r)\,\left(y(r) - \frac{\hbar(D-1)}{r}\right)\ =2M\left[\tilde{E}\ -\  \frac{1}{\gamma ^m}\hat{W}(\gamma\,r)\right]\ ,
\label{riccatikeyST}
\end{equation}
where $y(r)$ is defined as in (\ref{y}). With appropriate choice of variable
\begin{equation}
  w\ =\ \left(\frac{2M}{\hbar^2}\right)^{\frac{1}{m+2}}\,r\ ,
\label{wdef}
\end{equation}
with new unknown function and energy
\begin{equation}
\label{changesST}
   y(r)\ =\ (2M\,\hbar^m)^{\frac{1}{m+2}}\,\tilde{ \mathcal{Y}}(w) \ ,\qquad\quad
   \tilde{\veps}\ =\ \left(\frac{2M}{\hbar^2}\right)^{\frac{m}{m+2}}\,\tilde{E} \ ,
\end{equation}
the  equation (\ref{riccatikeyST}) becomes the RB Equation
\begin{equation}
\label{riccatiadiST}
   \pa_w\tilde{ \mathcal{Y}}(w)\ -\ \tilde{\mathcal{Y}}(w)\left(\tilde{\mathcal{Y}}(w)-\frac{D-1}{w}\right)\ =\ \tilde{\veps}(\tilde{\la})\ -\frac{1}{\tilde{\la}^m}\,\hat{W}\left(\tilde{\la} \,w\right)\ ,\qquad\quad \pa_w\equiv\frac{d}{dw}\ ,
\end{equation}
c.f. (\ref{riccati-bloch}) with different r.h.s., where
\begin{equation}
\label{effectiveST}
 \tilde{\la}\ =\ \left(\frac{\hbar^{2}}{2M}\right)^{\frac{1}{m+2}}\,\gamma \ ,
\end{equation}
see (\ref{1 over g}), plays a role of effective coupling constant, and $\tilde{\veps}(\tilde{\la})$ plays a role of energy.
Two remarks should be made: $(i)$ this equation does not contain explicit dependence on $\hbar$ and $M$;  $(ii)$ the parameter  $\tilde{\la}^{-1}$ is  the \textit{effective} {\it inverse} coupling constant, it appears instead of $\gamma^{-1}$. The latter implies that for  $\tilde{\mathcal{Y}}(w)$ and $\tilde{\veps}(\tilde\la)$ it should eventually be developed the expansions
\begin{equation}
\label{eq:YST}
\tilde{\mathcal{Y}}(w)\ =\  \sum_{n=0}^{\infty}\tilde{\mathcal{Y}}_n(w)\, \tilde{\la}^{-n}\ .
\end{equation}
and
\begin{equation}
\label{energyST}
   \tilde{\veps}({\tilde\la})\ =\  \sum_{n=0}^{\infty}\tilde{\veps}_n\, \tilde{\lambda}^{-n}\ ,
\end{equation}
respectively,
where $\tilde{\veps}_n$, $n=0,1,...$\,, are  the coefficients of the strong coupling expansion. It is evident that similar expansion (\ref{energyST}) occurs for the excited states.
Contrary to the weak coupling expansion it has been demonstrated explicitly that, at least,
for some particular cases of the potential (\ref{potential}) PT (\ref{energyST}) has finite radius
of convergence, see e.g. \cite{TurbinerStrong} and references therein. It is related with the fact
that the dominant term in the potential (\ref{potential}) is $r^m$ at $r \rar \infty$ while other
monomials are subdominant. Therefore, the phenomenon of the Dyson instability \cite{Dyson} does not occur.

It has to be emphasized that the equation (\ref{riccatiadiST}) has the same form as the Riccati equation (\ref{riccatigeneral}), this implies that all the formulas derived in Section \ref{PT} in the framework of  the Non-Linearization Procedure are valid and can be used to calculate $\tilde{\mathcal{Y}}_n(w)$ and $\tilde{\veps}_n$. Let us choose the potential (\ref{potentialp}) in the following form
\begin{eqnarray}
\label{selectionST}
 \la  &\ =\ &\ \left( \frac{2M}{\hbar^2} \right)^{\frac{1}{m+2}}  \gamma^{-1} \ ,\non\\
 V_k  &\ =\ &\ a_{m-k}\,w^{m-k}\quad\text{for}\quad k=0,\,1,\,...,\,m-2\ , \non\\
 V_k  &\ =\ &\ 0\ \quad\text{for}\quad k>m-2\ .
\end{eqnarray}
In this case, the unperturbed equation  (\ref{unperturbed}) is given by
\begin{equation}
   \pa_v\tilde{\mathcal{ Y}}_0(w)\ -\ \tilde{\mathcal{ Y}}_0(w) \left(\tilde{\mathcal{ Y}}_0(w) - \frac{D-1}{w} \right)\ =\ \tilde{\veps}_0\ -\ a_m\,w^m\ .
    \label{unperturbedST}
\end{equation}
If the solution of this equation is known, the corresponding square-integrable unperturbed wave function $\Psi_0$ is obtained using (\ref{PSI0}).
Then we use the  formulas (\ref{ycorrection}) and (\ref{ecorrection}) to construct PT.
As a result we obtain the  coefficients $\tilde{\veps}_k$ of the strong coupling expansion.
However, since the exact solution of (\ref{unperturbedST}) for $m > 2$ is unknown,  we need to find $\tilde{\mathcal{ Y}}_0(w)$ approximately. In the spirit of the present work, there are two possible ways to find it: $(i)$ one can choose a physically relevant trial function for the potential,
\begin{equation}
\label{W0}
   W_0(w)\ =\ a_{2p}\,w^{2p}\ ,\quad\quad\quad\ m\ =\ 2p\ ,
\end{equation}
and use the Non-Linearization Procedure assuming the convergence of PT; $(ii)$ we can construct a locally-accurate approximation of $\tilde{\mathcal{Y}}_0(w)$ for all $w \geq 0$.
Let us describe both approaches.

\subsubsection{Physically Relevant Trial Function}
\label{Physically}
Probably, the simplest physically relevant trial function, see \cite{TURBINER:1984} for discussion, for the potential (\ref{W0}) was proposed for the first time in \cite{TURBINER:1979,TURBINER:1981} handling the one-dimensional case. Undoubtedly, the similar trial function is also appropriate
for the $D$-dimensional case of the radial potentials,
\begin{equation}
\label{psi0}
   \Psi_{0,0}\ =\ e^{-a_{2p}^{1/2}\,{\frac{w^{p+1}}{p+1}}}\quad ,\quad
   \tilde{\mathcal{Y}}_{0,0}(w)\ =\ a_{2p}^{1/2}\,w^{p}\ ,
\end{equation}
where the second subscript marks the order of the correction to $\Psi_0$. By solving the \textit{inverse problem} one can find the potential $W_{0,0}(w)$ for which $\Psi_{0,0}$ is the exact solution,
\begin{equation}
\label{W00}
     W_{0,0}(w)\ =\ a_{2p}w^{2p}\ -\ a_{2p}^{1/2}(p+D-1) w^{p-1}\quad \mbox{with} \quad \tilde{\veps}_{0,0}\ =\ 0\ .
\end{equation}
Taking the difference between the  potentials (\ref{W0}) and (\ref{W00}),
\begin{equation}
\label{W01}
    W_{0,1}(w)\ =\ W_0(w)\ -\ W_{0,0}(w)\ =\ a_{2p}^{1/2}(p+D-1)\,w^{p-1}\ ,
\end{equation}
one develop PT as in (\ref{seriespE}) and (\ref{seriespy}) with $W_{0,1}(w)$ as the perturbative potential for $W_{0,0}(w)$.
In this manner, it allows us to calculate the zeroth order coefficient $\tilde{\veps}_0$
in strong coupling expansion (\ref{energyST}) iteratively, in the form of an expansion
\begin{equation}
\label{e01}
    \tilde{\veps}_0\ =\ \tilde{\veps}_{0,0}\ +\ \tilde{\veps}_{0,1}\ +\ \tilde{\veps}_{0,2}\ +\ \ldots\ .
\end{equation}
The first two coefficients can be found analytically,
\begin{align}
 \tilde{\veps}_{0,0}\ &=\ 0\ ,\\
 \tilde{\veps}_{0,1}\ &=\ a_{2p}^{\frac{1}{p+1}}(p+D-1)\left(\frac{p+1}{2}\right)^{\frac{p-1}{p+1}}\,
 \frac{\Gamma(\frac{D+p-1}{p+1})}{\Gamma(\frac{D}{p+1})} \ ,
\end{align}
while several next ones can be obtained numerically.
The function $\tilde{\mathcal{Y}}_0(w)$ is expanded in a similar way as in (\ref{e01}),
\begin{equation}
   \tilde{\mathcal{Y}}_0(w)\ =\ \tilde{\mathcal{Y}}_{0,0}(w)\ +\ \tilde{\mathcal{Y}}_{0,1}(w)\
   +\ \tilde{\mathcal{Y}}_{0,2}(w)\ +\ \ldots \ ,
\end{equation}
where $\tilde{\mathcal{Y}}_{0,0}(w)$ is given by (\ref{psi0}). Evidently, this expansion can be converted into the expansion of the phase,
\begin{equation}
   \Phi_0\ =\ \Phi_{0,0}\ +\ \Phi_{0,1}\ + \Phi_{0,2}\ +\ \ldots
\end{equation}
where $\Phi_{0,0}=a_{2p}^{1/2}\,\hbar\,{\dfrac{w^{p+1}}{p+1}}$. Using $\Psi_{0,0}$ as  entry one can also calculate explicitly  the first  approximation  to $\tilde{e}_1$ in the strong coupling expansion,
\begin{equation}
\label{e10}
\tilde{e}_{1,0} \ =\
   a_{m-1}\left(\frac{p+1}{2a_{2p}^{1/2}}\right)^{\frac{2p-1}{p+1}}\,\frac{\Gamma(\frac{D+2p-1}{p+1})}
          {\Gamma(\frac{D}{p+1})}\ .
\end{equation}
The next terms, e.g. $\tilde{e}_{1,1}$, $\tilde{e}_{1,2}$, $...$, can be only  written  in the form of nested integrals, similar to those as in expression (\ref{ecorrection}) and their computation is a numerical procedure. Of course, this procedure can be extended to calculate higher coefficients in the expansion (\ref{energyST}).

\subsubsection{Interpolation}

One can construct another physically relevant trial function for the RB Equation (\ref{unperturbedST}).
Let us find the asymptotic behavior of $y(r)$ at small $w$,
\[
y\ =\ (2M \hbar^m)^{\frac{1}{m+2}}\ \times
\]
\begin{equation}
\label{scalingrsmall}
\left(\frac{\tilde{\veps}_0}{D}\,w\ +\ \frac{\tilde{\veps}_0^{2}}{D^2\,(D+2)}w^3-\frac{a_{2p}\,\delta_{p,3/2}}{D+3}w^{4}\ +\ \frac{2\tilde{\veps}_0^3-a_{2p}D^3(D+2)\delta_{p,2}}{D^3(D+2)(D+4)}w^5+\  \ldots\right)
\end{equation}
and large $w$
\[
y\ =\ (2M \hbar^m)^{\frac{1}{m+2}}\ \times
\]
\begin{equation}
\label{scalingrlarge}
\left( a_{2p}^{1/2}\,w^{p} +\ \frac{1}{2}\,(D+p-1)\,w^{-1}\ -\ \frac{1}{2}\,\tilde{\veps}_0\,w^{-p}+\frac{4(D-1)(D-3)-m(m-4)}{32\,a_{2p}^{1/2}}\,w^{-p-2}+\ \ldots\right)\ .
\end{equation}
The first term in (\ref{scalingrlarge}) chosen as zero approximation corresponds to (\ref{psi0}). We have to note that large $w$ behavior can be obtained in two ways: taking either large $r$ at fixed $\hbar$, or \textit{small} $\hbar$ at fixed $r$; therefore the expansion (\ref{scalingrlarge}) should be valid in the semiclassical regime, see (\ref{wdef}). A more sophisticated  approximation for $y(r)$, and consequently for $\tilde{\mathcal{Y}}_{0,0}(w)$, is the result of the interpolation between expansions (\ref{scalingrsmall}) and (\ref{scalingrlarge}). The construction of such an interpolation is presented in the Section \ref{Approximant}. As for the cubic anharmonic oscillator, we show how this interpolation leads to a locally accurate approximation.

\subsection{The Strong Coupling Expansion from the Generalized Bloch Equation}

There is an interesting connection between the strong coupling expansion developed for the GB equation (\ref{Bloch}) and the expansion of $y(r)$ at small $r$, see (\ref{yrsmall}). For the case of polynomial potential of degree $m$ (\ref{potential}),
let us take the equation (\ref{riccatigeneral}), make change of variable (\ref{changeu}), $u=g r$,
then, following (\ref{r_to_r.g^m}) and (\ref{schroedingerST}),
make rescaling the energy
\begin{equation}
    E\ =\ \left(\frac{\hbar^2}{2M}\right)^{\frac{m}{m+2}}
    \ g^{2\frac{m-2}{m+2}}\ \tilde{\veps}\ ,
\end{equation}
and unknown function
\begin{equation}
   y(r)\ =\ \left(4 M^2 \hbar^{m-2} \right)^{\frac{1}{m+2}}\ g^{\frac{m-6}{m+2}}\
   \tilde{\mathcal{Z}}(u) \ .
\end{equation}
Finally, we arrive at
\begin{equation}
\label{blochstrong}
  \pa_u \tilde{\mathcal{Z}}(u)\ -\ \tilde{\mathcal{Z}}(u)\left(\tilde{\la}^{-2}\tilde{\mathcal{Z}}(u)-\frac{(D-1)}{u}\right)\ =\ \tilde{\veps}\ -\ \tilde{\la}^{-m}\,\hat{V}(\tilde{\la} u)
\end{equation}
c.f. (\ref{Bloch}), where the effective coupling constant $\tilde{\lambda}$ is given by (\ref{effectiveST}),
\[
 \tilde{\la}\ =\ \left(\frac{\hbar^{2}}{2M}\right)^{\frac{1}{m+2}}\,\gamma \ ,
\]
with
\[
   \gamma\ =\ g^{\frac{4}{m+2}}\ ,
\]
see (\ref{1 over g}).
This equation is the GB Equation for the strong coupling regime: after multiplication by $\tilde{\la}^2$ the l.h.s. is the same as in (\ref{Bloch}) - the GR equation for weak coupling regime - while the r.h.s. is different,
\begin{equation}
\label{blochstrong-final}
  \tilde{\la}^{2}\,\pa_u \tilde{\mathcal{Z}}(u)\ -\ \tilde{\mathcal{Z}}(u)\left(\tilde{\mathcal{Z}}(u)\ -\ \tilde{\la}^{2} \frac{(D-1)}{u}\right)\ =\ \tilde{\la}^{2}\,\tilde{\veps}\ -\ \tilde{\la}^{-m+2}\,\hat{V}(\tilde{\la} u)\ .
\end{equation}
Now we can develop PT in powers of inverse effective coupling constant $\tilde{\la}^{-1}$
\begin{equation*}
\label{energySTB}
 \tilde{\veps}({\tilde\la})\ =\ \sum_{n=0}^{\infty}\, \tilde{\veps}_n\, {\tilde\la}^{-n} ,
\end{equation*}
see (\ref{energyST}), and
\begin{equation}
\label{zSTB}
   \tilde{\mathcal{Z}}(u)\ =\ \sum_{n=0}^{\infty}\,
   \tilde{\mathcal{Z}}_n (u)\, \ {\tilde\la}^{-n}\ ,
\end{equation}
hence, making expansions at $\tilde{\lambda}=\infty$.
We assume the coefficients $\tilde{\veps}_n$ are known \textit{a priori}, in particular, from PT developed for the RB equation at strong coupling regime (\ref{riccatiadiST}): (\ref{eq:YST}) and (\ref{energyST}). Inserting these two series (\ref{energyST}) and (\ref{zSTB}) in (\ref{blochstrong-final}) one can determine iteratively corrections $\tilde{\mathcal{Z}}_n(u), n=0,1,2 \ldots$ by solving the same first order differential equation with different r.h.s.
\begin{equation}
 \pa_u\tilde{\mathcal{Z}}_i(u)\ +\ \frac{(D-1)}{u}\,\tilde{\mathcal{Z}}_i(u)\ =\
 \tilde{\veps}_i\ ,\qquad i=0,1\ ,
\label{eq:i=0,1}
\end{equation}
and
\begin{equation}
  \pa_u\tilde{\mathcal{Z}}_{n}(u)\ +\ \frac{(D-1)}{u}\,\tilde{\mathcal{Z}}_n(u)\ =\ \tilde{\veps}_n + \sum_{i=0}^{n-2}\tilde{\mathcal{Z}}_i(u)\,\tilde{\mathcal{Z}}_{n-i-2}(u)
  \ -\ \de_{m,n}\,\hat{V}(u)\ ,\ n=2,3,\ldots\ .
\label{eq:zetak}
\end{equation}
Imposing the boundary condition $\tilde{\mathcal{Z}}(0)=\tilde{\veps}/D$ to the original equation (\ref{blochstrong-final}), we can find the boundary condition for the $n$th correction,
\begin{equation}
 \pa_u\tilde{\mathcal{Z}}(u)\Bigr|_{u=0}\ =\ \frac{\tilde{\veps}_n}{D}\ .
\label{boundaryzeta}
\end{equation}
The equations (\ref{eq:i=0,1}) - (\ref{eq:zetak}) are solved in elementary way, in particular,
\begin{align}
 \tilde{\mathcal{Z}}_0(u)&\ =\ \frac{\tilde{\veps}_0}{D}u\ ,\non\\
 \tilde{\mathcal{Z}}_1(u)&\ =\ \frac{\tilde{\veps}_1}{D}u\ ,\non\\
 \tilde{\mathcal{Z}}_2(u)&\ =\ \frac{\tilde{\veps}_2}{D}u\ +\ \frac{\tilde{\veps}_0^2}{D^2(D+2)}u^3\ ,                 \non\\
 \tilde{\mathcal{Z}}_3(u)&\ =\ \frac{\tilde{\veps}_3}{D}u\ +\ \frac{2\,\tilde{\veps}_0\,\tilde{\veps}_1}{D^2(D+2)}u^3\ -\ \delta_{m,3}\,\sum_{k=2}^{m}\frac{a_k}{D+k}u^{k+1}\ ,\non\\
 \tilde{\mathcal{Z}}_4(u)&\ =\ \frac{\tilde{\veps}_4}{D}u\ +\ \frac{2\,\tilde{\veps}_0\,\tilde{\veps}_2+\tilde{\veps}_1^2}{D^2(D+2)}u^3\ +\ \frac{2\,\tilde{\veps}_0^3}{D^3(D+2)(D+4)}u^5\ -\ \de_{m,4}\,\sum_{k=2}^{m}\frac{a_k}{D+k}u^{k+1}\ ,\non
\end{align}
while the general $n$th correction is given by
\begin{equation}
    \tilde{\mathcal{Z}}_n(u) \ =\ \dfrac{\tilde{\veps}_n}{D}u\  +\ u^{1-D}\sum_{k=0}^{n-2}\int_{0}^{u}\tilde{\mathcal{Z}}_k(s)\,\tilde{\mathcal{Z}}_{n-k-2}(s)\,s^{D-1}\,ds\ -\ \delta_{m,n}\sum_{k=2}^{m} \dfrac{a_k\,u^{k+1}}{D+k} \ ,
\label{zstrong}
\end{equation}
at $n \geq 2$.
Note that it is not guaranteed the boundary condition $\tilde{\mathcal{Z}}(\infty)=+\infty$ is fulfilled for all these corrections. It implies that PT in inverse powers of
$\tilde{\la}$ for $\tilde{\mathcal{Z}}(u)$  has a finite radius of convergence in $u$.
Thus, the corrections $\tilde{\mathcal{Z}}_n(u)$ make sense in a bounded domain in $u$ only,
see discussion below.
From (\ref{zstrong}) one can see that any correction is a finite-degree polynomial in $u$,
\begin{equation}
   \tilde{\mathcal{Z}}_n(u)\ =\
\begin{cases}
 \displaystyle\sum_{k=0}^{[\frac{n}{2}]}\al_{2k}^{(n)}u^{2k+1}\ -\ \de_{m,n}\sum_{k=2}^{m} \dfrac{a_k\,u^{k+1}}{D+k} \quad \text{for\quad $n\leq m+1$}\ ,\\ \\
 \displaystyle \sum_{k=0}^{n}\alpha^{(n)}_ku^{k+1} \quad \text{for\quad $n> m+1$}\ ,
\end{cases}
\label{znstrong}
\end{equation}
where  $\al_k^{(n)}$ are real coefficients and the symbol $[\,\,]$ denotes the integer part.
It is easy to see that
\begin{equation}
\al_0^{(n)}=\frac{\tilde{\veps}_n}{D}\ ,
\end{equation}
following (\ref{boundaryzeta}), while the next coefficient always vanishes,
\begin{equation}
   \al_1^{(n)}=0\ ,
\label{eq:a1vanish}
\end{equation}
for any $n$. The coefficient $\al_{2}^{(n)}$ can be found in the form of finite sum
\begin{equation}
 \alpha_2^{(n)}=\frac{1}{D^2(D+2)}\sum _{k=0}^{n-2} \tilde{\veps}_k \tilde{\veps}_{n-k-2}\
\label{eq:a2} \ ,
\end{equation}
while the next coefficient again vanishes,
\begin{equation}
\al_3^{(n)}=0\ ,
\label{eq:a3vanish}
\end{equation}
for any $n$, c.f. (\ref{eq:a1vanish}).
From (\ref{znstrong}) one can see that $\tilde{\mathcal{Z}}_n(u)$ for $n < m$ and  $n = m+1$ is always an odd polynomial in $u$:
\begin{equation*}
 \tilde{\mathcal{Z}}_n(-u)\ =\ -\tilde{\mathcal{Z}}_n(u)\ ,
\end{equation*}
while for $n=m$ this property can be lost.
In the case of an even anharmonic potential, $V(r)=V(-r)$,
all odd corrections $\tilde{\mathcal{Z}}_{2n+1}(u)$ and $\tilde{\veps}_{2n+1}$
vanish. In turn, all even corrections $\tilde{\mathcal{Z}}_{2n}(u)$ are odd polynomials in $u$ of degree $(2n+1)$ if $2n > m+1$.

The perturbative approach, which was used to calculate the strong coupling expansion of
$\tilde{ \mathcal{Z}}(u)$, is related to the asymptotic behavior of $\tilde{ \mathcal{Z}}(u)$ at small $u$. Once we established the polynomial structure of the corrections $\tilde{\mathcal{Z}}_n(u)$, one can sum up some subseries of the perturbative expansion (\ref{zSTB}) keeping the degree of $u$ fixed. For example, the sum
\begin{equation}
\label{eq:subseries1}
  \left(\sum_{n=0}^{\infty}\al_{0}^{(n)}\tilde{\la}^{-n}\right)\, u\ =\ \frac{ \tilde{\veps}}{D}u\ ,
\end{equation}
corresponds to summing up all terms $O(u)$ in (\ref{zSTB}). The next sum is carried out
with all terms $O(u^3)$. Using (\ref{eq:a2}) one can show that it corresponds to
\begin{equation}
\label{eq:subseries2}
 \sum_{n=2}^{\infty}\al_{2}^{(n)}\tilde{\la}^{-n}u^3\ -\ \frac{ a_2\,\tilde{\la}^{-m}}{D+2}u^3\
 =\ \frac{ \tilde{\la}^{-2}\,\tilde{\veps}^2\ -\ a_2\,\tilde{\la}^{-m}D^2}{D^2(D+2)}u^3\ .
\end{equation}
The next sum contains coefficients in front of $u^4$ terms. Following (\ref{eq:a3vanish}), this sum has a single term,
\begin{equation}
\label{eq:subseries3}
   -\frac{a_3\,\tilde{\lambda}^{-m}}{D+3}u^4\ .
\end{equation}
This summation procedure can be extended to higher order terms $u^n, n > 4$. After performing such a summations we obtain the coefficients in front of all terms in the Taylor expansion in powers of $u$ for the function $\tilde{\mathcal{Z}}(u)$, and, eventually, for $y(r)$.
From another side one can construct the expansion in powers of $u$ directly from (\ref{blochstrong}),
\[
 \tilde{\mathcal{Z}}(u)\ = \frac{\tilde{\veps} }{D}u \ +\  \frac{ \tilde{\la}^{-2}\,\tilde{\veps}^2\ -\ a_2\,\tilde{\la}^{-m}D^2}{D^2(D+2)}u^3 \ -\
\frac{ a_3\,\tilde{\la}^{-m} }{D+3 }\,u^4\ -\
\]
\begin{equation}
\label{tZu}
 \frac{a_4\,\tilde{\la}^{-m}  D^3(D+2) -2\,\tilde{\veps}\,\tilde{\la}^{-2} (\tilde{\veps}^2\tilde{\la}^{-2}-a_2\,\tilde{\la}^{-m}D^2)}{D^3 (D+2) (D+4) }\,u^5
\ +\ \ldots \ .
\end{equation}

One can note the exact correspondence between the first three terms of the expansion (\ref{tZu}) and the subseries previously defined in (\ref{eq:subseries1}), (\ref{eq:subseries2}) and (\ref{eq:subseries3}). In this way we can see the connection between the strong coupling expansion, developed on the basis of the GB equation, and the small $u$ one: both expansions lead to the same representation of $\tilde{\mathcal{Z}}(u)$ and, eventually, of $y(r)$. Finally, this connection clarifies why the boundary condition $\tilde{\mathcal{Z}}(\infty)=+\infty$ can not always be fulfilled.

\section{The Approximant}
\label{Approximant}

Now we formulate a prescription for the  construction of a locally-accurate, uniform approximation of the ground state wave function. We denote this approximation by $\Psi_{0,0}^{(t)}$ and call it the \textit{Approximant}. Here the subscripts indicate explicitly the quantum numbers of the ground state $(n_r=0\,,\, l=0)$. It is convenient to follow the exponential representation of the wavefunction, i.e. $\Psi_{0,0}^{(t)}=e^{-\frac{1}{\hbar}\Phi_t}$, since, the phase $\Phi$ of the exponential representation is not that sharp as the wave function $\Psi$, being sufficiently smooth. Hence, we focus on the construction of the trial phase $\Phi_t$.

We follow the prescription that the phase $ \Phi_t$ interpolates between small and large distance $r$ expansions (\ref{phirsmall}) and (\ref{phirlarge}) and also the expansions of the strong coupling regime (\ref{scalingrsmall}) and (\ref{scalingrlarge}).
One can see immediately that the simplest interpolation based on that prescription is of the following form
\begin{equation}
  \frac{1}{\hbar}\Phi_t\ =\ \frac{\tilde{a}_0\ +\ \tilde{a}_1\,g\,r+\frac{1}{g^2}{\hat V}(r\,; \tilde{a}_2, \ldots , \tilde{a}_{m})}{\sqrt{\frac{1}{g^2\,r^{2}}{\hat V}(r\,;\,\tilde{b}_2, \ldots ,\tilde{b}_{m})}}\ +\
  \text{Logarithmic Terms\,($r\,;\,\{\tilde{c}\}$)}\ ,
\label{generalrecipe}
\end{equation}
cf. \cite{Turbiner2005} for $D=1$ at $m=4$. Without loss of generality we put $\tilde{b}_2=1$. Here ${\hat V}(r;\{\tilde a\})$ and ${\hat V}(r;\{\tilde b\})$ are modified potentials of the form of the original potential (\ref{Hamiltonian}), (\ref{potential}): instead of the external parameters $\{a\}$ some free parameters $\{\tilde{a}\}$ and $\{\tilde{b}\}$ are taken, respectively. Logarithmic terms can depend on free parameters $\{\tilde c\}$ as well. All those parameters will later be fixed by making the variational calculation with (\ref{generalrecipe}) taken as the phase of the trial function, see a discussion below. At $g =0$ the phase (\ref{generalrecipe}) becomes $a_0 + a_2 r^2$ that correspond to radial harmonic oscillator.
Formula (\ref{generalrecipe}) is the key result of this article.

The phase $\Phi_t$ has the outstanding property that with an appropriate choice of  parameters $(\tilde{a}_0,\tilde{a}_1; \tilde{a}_2,\tilde{b}_2,\ldots  \tilde{a}_k, \tilde{b}_k; \{\tilde{c}\})$ the generating function $G_0(r)$ is   partially (or exactly)  reproduced.
In particular, in order  to reproduce exactly the dominant term in expansion (\ref{phirlarge}) the condition
\begin{equation}
    \tilde{a}_m\ =\ \frac{2\,a_{m}^{1/2}}{m+2}\,\tilde{b}_{m}^{1/2}
\label{eq:consrinfy}
\end{equation}
is  imposed.
To mimic possible appearance of logarithmic terms in expansion (\ref{phirlarge}), in particular, in  $G_0(r)$ and $G_2(r)$, some logarithmic terms to phase $\Phi_t$ should be added, see (\ref{second}). These terms, mostly coming from $G_0(r)$ and $G_2(r)$, can be modified by replacing the concrete (numerical) coefficients by set $\{\tilde{c}\}$ of free parameters. For example, for a concrete two-term anharmonic oscillator potential
\begin{equation}
\label{two-term-potential}
  V(r)\ =\ r^2\ +\ g^{m-2}\,r^m\ ,
\end{equation}
the generating function $G_0(r)$
\begin{equation}
\label{eq:g0twoterm}
    \frac{G_0(r)}{(2M)^{1/2}}\ =\ \frac{2}{m+2}r^2\left\{\sqrt{1+(gr)^{m-2}}+\frac{m-2}{4}\, {}_2F_1\left(\frac{1}{2},\frac{2}{m-2};\frac{m}{m-2};-(gr)^{m-2}\right)\right\}\ ,
\end{equation}
contains logarithmic terms emerging in the hypergeometric function ${}_2 F_1$  \cite{BATEMAN}, while the generating function $G_2(r)$ contains logarithmic terms explicitly
\begin{equation}
 \frac{g^2\,G_2(r)}{(2M)^{1/2}}\ =\ \frac{1}{4}\,\log\left[1+(gr)^{m-2}\right]\ +\ \frac{D}{m+2}\,\log\left[1+\sqrt{1+(gr)^{m-2}}\right]\ ,
\label{eq:g2twoterm}
\end{equation}
see (\ref{semiclassicalcorrection}).
In this case logarithmic terms added in (\ref{generalrecipe}) should be a certain \textit{minimal} modification of the logarithmic terms that appear in $G_0(r)$ and $G_2(r)$. For instance, a minimal modification of $g^2G_2(r)/(2M)^{1/2}$ is of the form
\begin{equation}
\frac{1}{4}\,\log\left[1+\tilde{c}\,(gr)^{m-2}\right]\ +\ \frac{D}{m+2}\,\log\left[1+\sqrt{1+\tilde{c}\,(gr)^{m-2}}\right]\ ,
\end{equation}
where $\tilde{c}$ is a free parameter, if $\tilde{c}=1$ it coincides with (\ref{eq:g2twoterm}).

In general, some constraints on parameters have to be imposed to guarantee that $\Phi_t$ has
the structure of (\ref{phirsmall}). The remaining free parameters are fixed by giving them values of optimal parameters in a trial function used in variational calculus.
In order to realize it, we use the  Variational Method for the radial Hamiltonian (\ref{schroedingerR}) taking the Approximant $\Psi_{0,0}^{(t)}$ as a trial function.
It is worth noting that the accuracy of variational calculations can be estimated using the connection between Variational Method and PT \cite{TURBINER:1980},  it will be discussed
in next Subsection.

It is worth mentioning that an approximation for the wave functions of excited states can be constructed using as a building block the phase for the ground state $\Phi_t$. In $D=1$, the $n$th wave function is labeled by a single quantum number $n$, $n=0, 1,2,...$ . One can use an approximant of the form
\begin{equation}
  \Psi_n^{(t)}(r)\ =\ P_n(r)\,e^{-\Phi_t}\ ,
\label{eq:n}
\end{equation}
where $P_n$(r) is a polynomial of degree $n$ with  real roots: its coefficients can be used as variational parameters. In $D>1$, due to the separation of variables in spherical coordinates for the Schr\"odinger  equation (\ref{SE}), it is sufficient to approximate the radial part of the wave function.
One can label the eigenstates  by the pair of quantum numbers $(n_r,\ell)$ omitting the magnetic quantum number. In this case we use an approximant of the form
\begin{equation}
 \Psi_{n_r,\ell}^{(t)}(r)\ =\ r^{\ell}\,P^{(\ell)}_{n_r}(r^2)\,e^{-\Phi_t}\ ,
\label{eq:n,l}
\end{equation}
where $P^{(\ell)}_{n_r}$ is a polynomial of degree $n_r$ with positive roots with coeffcients as variational parameters. In both cases, (\ref{eq:n}) and (\ref{eq:n,l}), $\Phi_t$ has the same form of the phase, constructed for the ground state (\ref{generalrecipe}), but with a  different set of  parameters $\{\tilde{a}\},\{\tilde{b}\},\{\tilde{c}\}$.

Another way to fix the coefficients in $P_n(r)$ is to impose orthogonality constraints: for chosen $n$ $P_n(r)$ (\ref{eq:n}) should be orthogonal to all $k$ previously constructed functions, $k=0,1, ..., (n-1)$. It fixes all parameters in (\ref{eq:n}) while the remaining
free parameters are treated as variational.
For fixed angular momentum $\ell$ a similar procedure is used for $P^{(\ell)}_{n_r}(r^2)$ and its parameters,  the orthogonality constraint is imposed in such a way that the eigenstates  with $k_r=0,1,\ldots (n_r-1)$ are orthogonal to the state with quantum number $n_r$. Again, the remaining free parameters are used as variational ones.

\subsection{Variational Method and PT}

The realization of the Non-Linearization Procedure does not require the knowledge of the whole spectra of the unperturbed problem. This property gives a freedom to  choose the unperturbed potential $V_0$ of  equation  (\ref{unperturbed}) by solving the \textit{inverse problem}. Exploiting this property we are able to find a relationship between PT and variational method \cite{TURBINER:1980} (for discussion \cite{TURBINER:1984}).

Let us consider square-integrable nodeless trial function $\Psi^{(t)}(r)$ at $r \in [0, \infty)$, thus it may correspond to a ground state, such that
\begin{equation}
\int_0^\infty (\Psi^{(t)})^2\,r^{D-1}\,dr\ <\ \infty\ .
\end{equation}
Notice that $\Psi^{(t)}$ is an eigenfunction of the operator (\ref{radialop}) with potential
\begin{equation}
          V_t\ =\ E_0\ +\
  \frac{\hbar^2}{2M}\ \frac{\pa_r^2\Psi^{(t)}+\frac{D-1}{r}\,\pa_r\Psi^{(t)}}{\Psi^{(t)}}\ .
\label{potential-t}
\end{equation}
We can always represent the original potential $V$ as the sum
\begin{equation}
   V\ =\ V_t\ +\ \la\,(V-V_t)\ \equiv\ V_0\ +\ \la\,V_1\ ,
\label{rewritten}
\end{equation}
where formal parameter $\la$ is placed to one, $\la=1$. One can consider a formal PT in powers of $\lambda$ and eventually set $\lambda=1$. In this context, the variational energy for the radial operator $\hat{h}_r$ (\ref{radialop}) is given by
\begin{align}
  E_{var}\ =\ &\frac{\int \Psi^{(t)}\,\hat{h}_r\,\Psi^{(t)}\, r^{D-1}\,dr}{\int (\Psi^{(t)})^2\, r^{D-1}\,dr}\ =\ E_0\ +\ \frac{\int \Psi^{(t)}\,V_1\,\Psi^{(t)}\, r^{D-1}dr}{\int (\Psi^{(t)})^2\, r^{D-1}dr}\\
  \ =&\ E_0\ +\ E_1\ \ .
\label{VariationalPerturbation}
\end{align}
This result suggests to interpret the variational energy as the first two terms in a perturbation theory where $(V-V_t)$ plays a role of perturbation potential. If the function $\Psi^{(t)}$ depends on certain parameters, we have a parameter-dependent variational energy, which can be minimized with respect to them. The variational principle guarantees that $E_{var}$ is above the ground state energy. The parameters that minimize $E_{var}$ are called optimal parameters. Calculating the next terms $E_2$, $E_3$, ... in the expansion (\ref{VariationalPerturbation}) using formulas (\ref{ycorrection}) and (\ref{ecorrection}), one can estimate the accuracy of the variational calculation and perturbatively improve it, if it is convergent. Choosing different trial function $\Psi_t$, we define different approximations for the energy $E$.

Taking partial sums we define approximations of different orders. In particular, the variational energy
\begin{equation}
E_0^{(1)}\ =\ E_0\ +\ E_1\ ,
\end{equation}
and
\begin{equation}
E_0^{(2)}\ =\ E_0\ +\ E_1\ +\ E_2\ ,
\end{equation}
correspond to first and second order approximations, respectively.
In general, the partial sum
\begin{equation}
E_0^{(n)}\ =\ E_0\ +\ E_1\ +\ \ldots\ +\ E_n\  .
\label{eq:ptvariational}
\end{equation}
defines the $n$th approximation. If we calculate corrections  for  $y_0=\Phi_t'$ in the framework of the Non-Linearization procedure, we can estimate  the accuracy of $\Phi_t$ and ultimately the accuracy of the Approximant used as trial function. Moreover, if the correction $y_1$ is bounded we know that (\ref{eq:ptvariational}) converges as $n \rar \infty$ \cite{TURBINER:1984}. Needless to say, this connection between the Variational Method and PT is also valid for excited states \cite{TURBINER:1980}.


\section{Cubic Anharmonic Oscillator}
\label{cubicAHO}

The simplest radial anharmonic oscillator is characterized by a cubic anharmonicity,
\begin{equation}
 V(r)\ =\ r^2\ +\ g\,r^3\ ,
\label{cubicpotential}
\end{equation}
c.f. (\ref{Hamiltonian}), (\ref{two-term-potential}) at $m=3$.
This potential is the subject of the forthcoming Section. It is worth noting before to proceeding to presentation that many features that the cubic anharmonic potential exhibits are also present in the general potential $V(r)\neq V(-r)$.

\subsection{PT in the Weak Coupling Regime}

For the cubic anharmonic oscillator the perturbative expansions of $\veps$ and $\mathcal{Y}(v)$
(\ref{seriespy}), (\ref{seriespE}) remain the same functionally as for the general oscillator,
\begin{equation*}
 \veps\ =\ \veps_0\ +\ \veps_1\,\la\ +\ \veps_2\,\la^2\ +\ \ldots\
\end{equation*}
with $\veps_0 = D$ and
\begin{equation*}
  \mathcal{Y}(v)\ =\ \mathcal{Y}_0\ +\  \mathcal{Y}_1\,\la\ +\ \mathcal{Y}_2\,\la^2\ +\ \ldots \ ,
\end{equation*}
respectively, see (\ref{changev}), (\ref{changes}). If it is assumed $2M=\hbar=1$, then $\la=g$, $v=r$ and $\veps=E$, $\mathcal{Y}=y$.
Since  potential (\ref{cubicpotential}) contains odd degree monomial $r^3$, the correction $\mathcal{Y}_n(v)$ is characterized by the infinite series (\ref{small}) and (\ref{large})
at small and large $v$, respectively.
The first three terms in expansion of $\mathcal{Y}_n(v)$ at $n=1,2,3$ at both $v \rar 0$ and
$v \rar \infty$ asymptotics are presented in Appendix \ref{appendix:A}.
The first corrections $\veps_1$ and $\mathcal{Y}_1(v)$ can be calculated in closed analytic form using (\ref{ycorrection}) and (\ref{ecorrection}),
\begin{equation}
 \veps_1\ =\ \frac{\Gamma(\frac{D+3}{2})}{\Gamma(\frac{D}{2})}\quad ,\quad
 \mathcal{Y}_1(v)\ =\ \dfrac{e^{v^2}}{2\,v^{D-1}}\left\{\veps_1\,\gamma \left(\frac{D}{2},v^2\right)-\gamma\left(\dfrac{D+3}{2},v^2\right)\right\}\ ,
\end{equation}
where $\Gamma(a)$ and $\gamma(a,b)$ denote the (in)complete gamma functions,
respectively, see \cite{BATEMAN}. Note that the higher energy corrections $\veps_n$ can be computed only numerically. However, the behavior of $\veps_n$ at large $D$ can be obtained in $1/D$-expansion analytically. This expansion can be constructed via the Non-Linearization Procedure, it is not be presented in this paper.

\subsection{Generating Functions}

We can determine the coefficients of $\mathcal{Y}_{n}(v)$, i.e. $c_{k}^{(n)}$ in (\ref{large}),  by algebraic  means. In general, they  are written in terms of the  corrections $\veps_0$, $\veps_{1}$, $\ldots$, $\veps_n$, see Appendix \ref{appendix:A} for explicit formulas for $n=1,2,3$.

It was pointed out in \cite{DOLGOVPOPOV1979} as well as in \cite{DOLGOVPOPOV1978}, see also
\cite{TURBINER:1980,TURBINER:1984}, that the general expression for several first coefficients $c_0^{(n)}$, $c_{2}^{(n)}, \ldots$  can be obtained by solving recurrence relations. For example, $c_0^{(n)}$ satisfies non-linear recurrence relation
\begin{equation}
 c^{(n)}_0\ = \ -\frac{1}{2}\sum_{k=1}^{n-1}c_0^{(k)}c_0^{(n-k)}\ ,\quad\quad\quad c_0^{(1)}\ = \  \frac{1}{2} \  ,
 \label{recurrence1}
\end{equation}
as for $c_{2}^{(n)}$ the recurrence relation is linear
\begin{equation}
 c_{2}^{(n)}\ =\ \frac{1}{2}\,(2n+D)\,c_0^{(n)}\ -\ \sum_{k=1}^{n-1}c_0^{(k)}c_{2}^{(n-k)}\ ,\quad\quad\quad c_2^{(1)}\ =\ \frac{1}{4}(D+2)\ ,
\label{recurrence2}
\end{equation}
the solution of  (\ref{recurrence2}) requires the  knowledge of the coefficients $c_0^{(n)}$.
It is evident that in order to find $c_{3}^{(n)}$, the coefficients $c_0^{(n)}$, $c_{2}^{(n)}$ are needed, recurrence relation remains linear, etc. The simplest way to solve these recurrence relations is by using the generating functions. Surprisingly, we can calculate these generating functions straightforwardly via the algebraic, iterative procedure derived from the GB equation, see (\ref{BlochSol}): we demonstrated that $k$th generating function is nothing but $\mathcal{Z}_k(u)$!
For example, the first two generating functions are
\begin{align}
 \mathcal{Z}_0(u)&\ =\ u\, \sqrt {1+u}
\label{eq:Y0cubic}\ ,\\
 \mathcal{Z}_2(u)&\ =\  \frac{u+2D\left(1+u-\sqrt{1+u}\right)}{4u(1+u)}\ ,
\end{align}
where $u=(gr)$. From these expressions, the explicit solutions of the equations (\ref{recurrence1}) and (\ref{recurrence2}) can be derived,
\begin{equation}
 c_0^{(n)}\ =\ \frac{(-1)^{n+1}\,\Gamma(2n+1)}{2^{2n-1}\Gamma(n)\,\Gamma(n+1)}\quad , \quad
 c_{2}^{(n)}\ =\ \frac{(-1)^{n+1}}{2}\left(1+D\frac{\Gamma(2n+1)}{2^{2n}\,\Gamma(n+1)^2}\right)
 \ .
\label{solutions}
\end{equation}
Note that the coefficients $c_0^{(n)}$ are $D$-independent, $c_{2}^{(n)}$ depends on $D$ linearly. It can be demonstrated by induction that $c_{k}^{(n)}$ is polynomial in $D$ of degree $(k-1)$.

As mentioned before, an important property  of  the generating functions $\mathcal{Z}_0(gr)$, $\mathcal{Z}_2(gr)$, $\mathcal{Z}_3(gr)$, ... is related to the asymptotic behavior of the function $y$ at large $v$. From  (\ref{asymptoticgeneral}) we have
\[
  y\ =\ (2M \hbar^2)^{\frac{1}{4}}\ \times
\]
\begin{equation}
  \left( \la^{1/2}v^{3/2}\ +\ \frac{1}{2\la^{1/2}}v^{1/2}\ -\ \frac{1}{8\la^{3/2}}v^{-1/2}\ +\
  \frac{2 D+1}{4}v^{-1}\ -\ \frac{(8\la^2\veps-1)}{16\la^{5/2}}v^{-3/2}\ +\ \ldots\right)\ ,
\label{inftycubic}
\end{equation}
see (\ref{changev}). Note that the first four terms of this expansion are independent of $\veps$. The expansion of $(2M)^{1/2}\mathcal{Z}_0(gr)$ at large $r$ can be transformed into an expansion
for large $v$, namely
\begin{equation}
 (2M)^{1/2}\mathcal{Z}_0\ =\ (2M \hbar^2)^{\frac{1}{4}}\,\left(\la^{1/2}v^{3/2}\ +\ \frac{1}{2\la^{1/2}}v^{1/2}\ -\ \frac{1}{8\la^{3/2}}v^{-1/2}\ +\ \frac{1}{16 \la^{5/2}}v^{-3/2}\ +\ \ldots\right)\ .
\end{equation}
It reproduces exactly the expansion (\ref{inftycubic}) up to  $O(v^{-1/2})$.
At large $r$, the function $(2M)^{1/2}\la^2\mathcal{Z}_2(gr)$ contributes to the expansion (\ref{inftycubic}) in a similar way starting from  $O(v^{-1})$,
since
\begin{equation}
 (2M)^{1/2}\la^2\mathcal{Z}_2\ = \ (2M \hbar^2)^{\frac{1}{4}}\,
 \left(\frac{2 D+1}{4}v^{-1}\ -\ \frac{D}{2 \la^{1/2}}v^{-3/2}\ -\ \frac{1}{4 \la}v^{-2}\
 +\ \ldots\right)\ .
\end{equation}
Note that in the expansion of $(2M)^{1/2}\mathcal{Z}_0(gr)$ plus $(2M)^{1/2}\la^2\mathcal{Z}_2(gr)$ at large $v$ reproduces exactly the coefficient in front of   $O(v^{-1})$ in expansion (\ref{inftycubic}). The expansion of next function $(2M)^{1/2}\la^3\mathcal{Z}_3$ starts from  $O(v^{-3/2})$. Together with the first two
$\mathcal{Z}_{0,2}$ it reproduces the coefficient in front of $O(v^{-3/2})$ term exactly.

\subsection{ The Approximant and Variational Calculations }

The first two generating functions, $G_0(r)$ and $G_2(r)$ in expansion (\ref{semiclassical})
are given by
\begin{align}
 G_0(r)&\ = \ \frac{(2M)^{1/2}}{g^2}\left(\frac{2 (-2+3 g r)(1+gr)^{3/2} }{15 }\right)\ ,\\
 G_2(r)&\ = \ \frac{(2M)^{1/2}}{g^2}\left(\frac{1}{4}\log[1+gr]+D\log\left[1+\sqrt{1+gr}\right]\right)\ .
\label{eq:cubicG2}
\end{align}
The next two functions $G_3(r)$ and $G_4(r)$ are presented explicitly in Appendix \ref{appendix:A}. The function $G_0(r)$ contains no logarithmic terms.

The generating functions $G_0(r)$ and $G_2(r)$ serve the building blocks for the construction
of the Approximant. Following the general prescription (\ref{generalrecipe}) we write the Approximant in the exponential representation $\Psi_{0,0}^{(t)}=e^{-\Phi_t}$ with
\begin{equation}
 \Phi_t \ =\
 \frac{\tilde{a}_0\ +\ \tilde{a}_1\,gr+\tilde{a}_2\,r^2\ +\ \tilde{a}_3\,g\,r^3}
 {\sqrt{1\ +\ \tilde{b}_3\, g\, r}}\ +\ \frac{1}{4}\,\log[1\ + \tilde{b}_3\, g\,r] + D \log \left[1\ +\ \sqrt{1\ +\ \tilde{b}_3\, g\,r}\right]\ ,
\label{trialcubic}
\end{equation}
where we set
\begin{equation}
\label{a3}
\tilde{a}_3\ =\ \dfrac{2}{5}\,\tilde{b}_3^{1/2}\ ,
\end{equation}
in order to reproduce exactly the leading term in asymptotic behavior of phase at $r \rar \infty$.
Here, the logarithmic terms in $\Phi_t$ represent a \textit{minimal} modification of ones in $G_2(r)$, cf. (\ref{eq:cubicG2}). Additionally, we impose a constraint
\begin{equation}
\label{a1}
 \tilde{a}_1 \ =\ \frac{ \tilde{b}_3 }{4}\,(2\,\tilde{a}_0\ - D\ -1)\ ,
\end{equation}
in order to guarantee vanishing $\Phi_t^{\prime}$ at $r=0$, thus, no linear term in expansion of (\ref{trialcubic}). Eventually, the Approximant in its final form depends on three free parameters $\{\tilde{a}_0, \tilde{a}_2, \tilde{b}_3\}$ {\it only}, it reads
\begin{equation}
\label{trialcubic-psi}
 \Psi_{0,0}^{(t)}\ =\ \frac{1}{\left(1\ +\ \tilde{b}_3\,g\,r\right)^{1/4}\left(1\ +\ \sqrt{1\ +\ \tilde{b}_3 \,g\,r}\right)^D}\,
 \exp\left(-\frac{\tilde{a}_0\ +\ \tilde{a}_1\,g\,r\ +\ \tilde{a}_2\,r^2+\tilde{a}_3\,g\,r^3}{\sqrt{1\ +\ \tilde{b}_3\, g\, r}}\right)\ ,
\end{equation}
where $\tilde a_{3,1}$ are defined by constraints (\ref{a3}), (\ref{a1}).
This is the eventual expression for the approximate, 3-parametric ground state wave function,
{\color{blue} which is key result of Section on cubic radial anharmonic oscillator}.
As the first step this function is used as a trial function in variational calculation in order to fix 3 free parameters.

As for excited states at $D>1$ $(n_r,\ell)$ the Approximant has the form
\begin{equation}
\label{trialcubic-psi-excited}
 \Psi_{n_r,\ell}^{(t)}\ =\ \frac{ r^{\ell} P^{(\ell)}_{n_r}(r^2)}{\left(1\ +\ \tilde{b}_3\,g\,r\right)^{1/4}\left(1\ +\ \sqrt{1\ +\ \tilde{b}_3\, g\,r}\right)^D}\,
 \exp\left(-\ \frac{\tilde{a}_0\ +\ \tilde{a}_1\,g\,r\ +\ \tilde{a}_2\,r^2\ +\ \tilde{a}_3\,g\,r^3}{\sqrt{1\ +\ \tilde{b}_3\, g\, r}}\right)\ ,
\end{equation}
where $P^{(\ell)}_{n_r}(r^2)$ is the polynomial of degree $n_r$ with $n_r$ real positive roots and with leading term $r^{2\,n_r}$. $n_r$ coefficients of $P^{(\ell)}_{n_r}(r^2)$ are found by imposing the $n_r$ orthogonality conditions:
$$ (\Psi_{n_r,\ell}^{(t)},\Psi_{k,\ell}^{(t)})\ =\ 0\ ,\ k=0,\ldots (n_r-1)\ .$$
After imposing the orthogonality conditions and constraints (\ref{a3}), (\ref{a1}) on $a_{3,1}$,
the function (\ref{trialcubic-psi-excited}) depends eventually on 3 free parameters $\{\tilde{a}_0, \tilde{a}_2, \tilde{b}_3\}$, which are found by making minimization of the variational energy.
For all studied states the optimal parameters $\{\tilde{a}_0, \tilde{a}_2, \tilde{b}_3\}$ demonstrate smooth behavior versus $g$ and $D$.

Variational energy calculation with trial function $\Psi_{0,0}^{(t)}$ (\ref{trialcubic-psi}) requires to perform a numerical integration of the integrals in numerator and denominator (\ref{VariationalPerturbation}), and also a numerical minimization. Computational code was written in FORTRAN 90 with use of the integration routine D01FCF from the NAG-LIB, which was built using the algorithm described in \cite{INT}. The optimization routine to find the variational parameters was performed using the program MINUIT of CERN-LIB. Let us mention that the use of Approximant $\Psi_{n_r,\ell}^{(t)}$ for excited states as trial functions in the form (\ref{trialcubic-psi-excited}), c.f. (\ref{eq:n}) or (\ref{eq:n,l}), also require numerical integration and minimization.

\begin{figure}[h]
	\centering
	\begin{subfigure}[t]{0.47\textwidth}
		\centering
		\includegraphics[width=\linewidth]{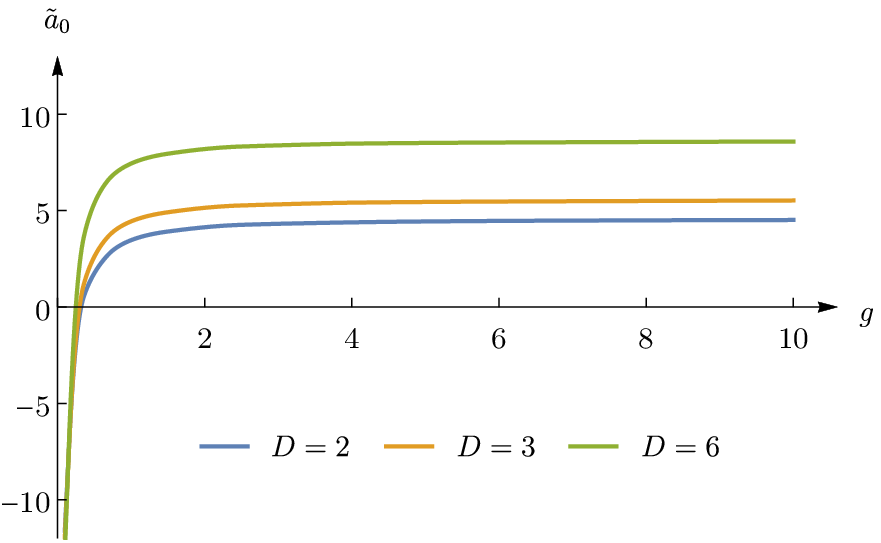} 
		\caption{} \label{fig:a0}
	\end{subfigure}
	\hfill
	\begin{subfigure}[t]{0.47\textwidth}
		\centering
		\includegraphics[width=\linewidth]{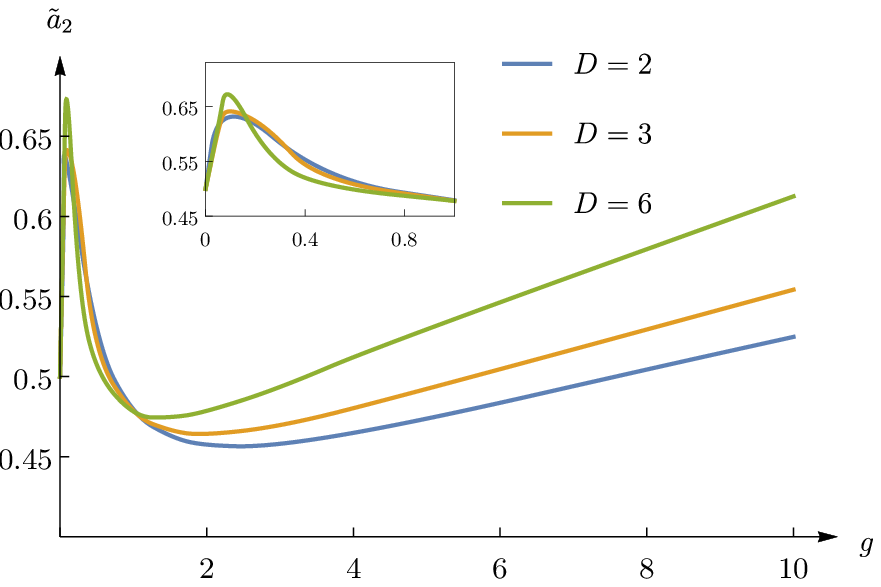} 
		\caption{} \label{fig:a3}
	\end{subfigure}
	
	\vspace{1cm}
	\begin{subfigure}[t]{0.5\textwidth}
		\centering
		\includegraphics[width=\linewidth]{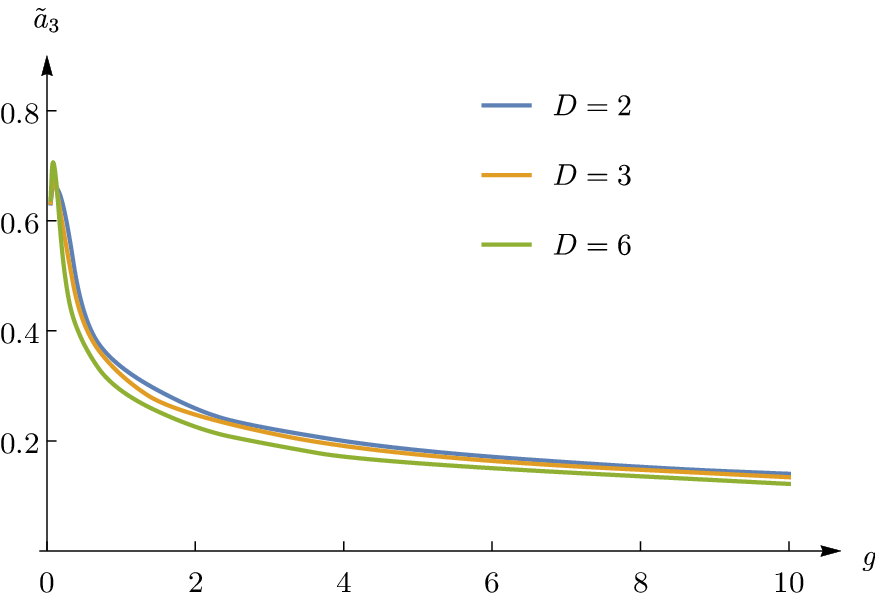} 
		\caption{} \label{fig:a2}
	\end{subfigure}
	\caption{Variational parameters  ${\tilde a}_0$, ${\tilde a}_2$ and ${\tilde a}_3$  of the ground state as functions of $g$ for $D=2,3,6$. }
	\label{fig:varpar}
\end{figure}

Without loss of generality we set $\hbar=1$ and $M=1/2$, thus, putting $v=r$, $\veps=E$, $\la=g$ and $\mathcal{Y}=y$, see (\ref{changev}), (\ref{changes}) and (\ref{effective}).
The calculations of variational energy for four low-lying states with quantum numbers
(0,0), (1,0), (0,1) and (0,2) for different values of $D$ and $g$ are presented in
Tables \ref{cubicres1} - \ref{cubicres3}. {\color{red} \bf In general, variational parameters are smooth, slow changing functions in $g$ without strong $D$ dependence. As example,
for the ground state (0,0) the plots of the parameters ${\tilde a}_{0,2,3}$ {\it vs} $g$ for $D=2,3,6$ are shown in Fig. \ref{fig:varpar}.}

For some states the variational energy $E_{var}=E_0^{(1)}$, the first correction $E_2$
to it as well as its corrected value $E_0^{(2)}=E_{var}+E_2$ are shown.
All states are studied for different values of dimension $D$ and coupling constant $g$.
For all cases the variational energy $E_0^{(1)}$ is obtained with absolute accuracy $10^{-7}-10^{-8}$ (7-8 s.d.).
This accuracy is found by calculating the second correction $E_2$ (the first correction to the variational energy). Furthermore, making comparison of energies $E_0^{(2)}$ with ones coming from the Lagrange Mesh method (see below) one can see that the accuracy 8 d.d. is reached.
This accuracy is confirmed independently by calculating the second correction $E_3$ to variational energy: this correction is $\lesssim 10^{-9}$ for any $D$ and $g$ we have studied. It indicates to very fast rate of convergence in Non-Linearization Procedure with trial function (\ref{trialcubic-psi}) taken as zero approximation.

To check the energies obtained in Non-Linearization Procedure we calculated them in the Lagrange Mesh Method \cite{BAYE}. It is known that the Lagrange Mesh Method in the form proposed and developed by D.~Baye with co-authors, see \cite{BAYE} and references therein, for various problems of atomic and molecular physics, provides easily 12 - 13 s.d. in energies. In present work we apply this method to study cubic radial anharmonic oscillator in different dimensions $D$. These studies are performed for one-dimensional case $D=1$ using the Hermite mesh, while for larger dimensions $D>1$ the Laguerre mesh is used. In both occasions 50 mesh points are used as maximum number of points. Details will be published elsewhere.

These numerical calculations show that all digits presented for $E_0^{(2)}$ - the variational energy with first (second order in PT) correction $E_2$ taken into account - in Tables \ref{cubicres1} - \ref{cubicres3} are exact(!): we obtain not less than nine decimal digits correctly in Non-Linearization Procedure. In general, the energies grow with increase of $D$ and/or $g$.
Classifying states by radial quantum number $n_r$ and angular momentum $\ell$ as $(n_r, \ell)$,
we mention the hierarchy of eigenstates which holds for any fixed integer $D$ and $g$: (0,0), (0,1), (0,2), (1,0).

It has to be noted that to the best of the authors knowledge these results represent the first accurate variational calculations (and corrections to them found in the Non-Linearization Procedure) of the energy of the low-lying states for the $D$-dimensional cubic anharmonic oscillator.

\begin{table}[ht]
 	\centering
 	\caption{Ground state energy for the cubic potential $(r^2\,+\,g\,r^3)$ for $D=1,2,3,6$
     and $g=0.1, 1, 10$, labeled by quantum numbers (0,0) for $D>1$ . Variational energy $E_0^{(1)}$, the first correction $E_2$ (rounded to three s.d.) found with use of $\Psi_{0,0}^{(t)}$, see text, the corrected energy $E_0^{(2)}=E_0^{(1)}+E_2$ shown. $E_0^{(2)}$ coincides with Lagrange Mesh results (see text) in nine displayed decimal digits, hence all printed digits are exact. }
 	\label{cubicres1}
 	\begin{tabular}{|c|c c c|ccc|}
\hline
   \multirow{2}{*}{$\quad\,g\quad\,$} & \multicolumn{3}{c|}{$D=1$} & \multicolumn{3}{c|}{$D=2$}
   \\
\cline{2-7}
 	& $\quad\quad\,  E_0^{(1)} \quad\quad\,$ & $\quad\quad\, -E_2       \quad\quad\,$
    & $\quad\quad\,  E_0^{(2)} \quad\quad\,$ & $\quad\quad\,  E_0^{(1)} \quad\quad\,$                                           & $\quad\quad\, -E_2       \quad\quad\,$ & $\quad\quad\,  E_0^{(2)} \quad\quad\,$
    \\
\hline
 	\rule{0pt}{5ex}0.1 & 1.053120300\quad & $5.39 \times 10^{-7}$ & 1.053119761 & 2.124027648
                       & $4.40\times10^{-7}$ & 2.124027208
    \\
 	1.0                & 1.387428891      & $4.00\times10^{-8}$   & 1.387428851 & 2.877490906                                 & $3.76\times10^{-8}$ & 2.877490868
    \\
 	10.0               & 2.729533139      & $6.56\times10^{-7}$   & 2.729532483 & 5.794213459                                 & $5.58\times10^{-7}$ & 5.794212901
    \\[8pt]
\hline
\hline
 	\multirow{2}{*}{$g$} & \multicolumn{3}{c|}{$D=3$} & \multicolumn{3}{c|}{$D=6$}
    \\
\cline{2-7}
 	& $E_0^{(1)}$        & $-E_2$  & $E_0^{(2)}$  & $E_0^{(1)}$  & $-E_2$ & $E_0^{(2)}$
    \\
\hline
 	\rule{0pt}{5ex}0.1  & 3.208922743 & $4.00\times10^{-7}$  & 3.208922343 & 6.528432540
    &  $2.02\times 10^{-7}$ & 6.528432338
    \\
 	1.0                 & 4.442965260 & $3.15\times10^{-8}$  & 4.442965229 & 9.465319951
 &  $1.85\times 10^{-8}$    & 9.465319933
    \\
 	10.0                & 9.094985589 & $4.23\times10^{-7}$  & 9.094985166 & 19.981458504                         &  $1.96\times 10^{-7}$    & 19.981458308
    \\[8pt]
\hline 		
 	\end{tabular}
\end{table}

\begin{table}[h]
 	\caption{The first excited state energy for the cubic potential $(r^2\,+\,g\,r^3)$
     for different $D$ and $g$. They labeled by quantum numbers (0,1) for $D>1$,
     as for $D=1$ it corresponds to 1st excited state.
     Variational energy $E_0^{(1)}$, the first correction $E_2$ found with use of
     $\Psi_{1,0}^{(t)}$ for $D=1$ and $\Psi_{0,1}^{(t)}$ for $D>1$, see text, the corrected energy $E_0^{(2)}=E_0^{(1)}+E_2$ shown, the correction $E_2$ rounded to three s.d.
     All printed digits for $E_0^{(2)}$ are exact.}
 	\centering
 	\label{cubicres2}
 	\begin{tabular}{|c|ccc|ccc|}
 		\hline
 		\multirow{2}{*}{$\quad\,g\quad\,$} & \multicolumn{3}{c|}{$D=1$} & \multicolumn{3}{c|}{$D=2$} \\ \cline{2-7}
 		& $\quad\quad\, E_0^{(1)}\quad\quad\,$ & $\quad\quad\, -E_2\quad\quad\,$ & $\quad\quad\,
   E_0^{(2)} \quad\quad\,$  & $E_0^{(1)}$ & $\quad\quad\, -E_2 \quad\quad\,$ & $\quad\quad\, E_0^{(2)} \quad\quad\,$
   \\
\hline
 	\rule{0pt}{5ex}0.1 & 3.208922765 & $4.21\times10^{-7}$ & 3.208922343 & 4.305557665 & $3.55\times10^{-7}$       & 4.305557309
   \\
 	1.0 & 4.442965265 & $3.59 \times 10^{-8}$ & 4.442965229 & 6.068723537 & $2.92 \times 10^{-8}$ & 6.068723507
   \\
   10.0 & 9.094985630 & $4.64\times10^{-7}$   & 9.094985166 & 12.579594377 & $3.48\times10^{-7}$ & 12.579594029       \\[8pt]
\hline\hline
 	\multirow{2}{*}{$g$} & \multicolumn{3}{c|}{$D=3$} & \multicolumn{3}{c|}{$D=6$}
   \\
\cline{2-7}
 	& $E_0^{(1)}$ & $-E_2$ & $E_0^{(2)}$ & $E_0^{(1)}$ & $-E_2$ & $E_0^{(2)}$
   \\
\hline
 	\rule{0pt}{5ex}0.1 & 5.412425220 & $2.86 \times 10^{-7}$ & 5.412424933 & 8.784695351 & $1.21 \times 10^{-7}$ & 8.784695230
   \\
 	1.0 & 7.745092165 & $2.41 \times 10^{-8}$ & 7.745092141 & 13.018486318 & $1.49 \times 10^{-8}$ & 13.018486303
   \\
   10.0 & 16.215748127 & $2.66 \times 10^{-7}$ & 16.215747861 & 27.841430199 & $1.37 \times 10^{-7}$ & 27.841430061
   \\[8pt]
\hline
 	\end{tabular}
\end{table}

\begin{table}[h]
 	\centering
 	\caption{The second excited state energy for the cubic potential $r^2\,+\,g\,r^3$ for different
    $D$ and $g$, they labeled by quantum numbers (0,2) for $D>1$, as for $D=1$ it corresponds
    to 2nd excited state $n_r=2$. Variational energy $E_0^{(1)}$ found with use of $\Psi_{2,0}^{(t)}$
    for $D=1$ and $\Psi_{0,2}^{(t)}$ for $D>1$, see text,
    the first correction $E_2$ and the corrected energy $E_0^{(2)}=E_0^{(1)}+E_2$ shown.
    Correction $E_2$ rounded to 3 s.d. All 9 printed decimal digits in $E_0^{(2)}$ are exact.}
 	\label{cubic2}
 	\begin{tabular}{|c|ccc|ccc|}
 		\hline
 		\multirow{2}{*}{$\quad\,g\quad\,$} & \multicolumn{3}{c|}{$D=1$} & \multicolumn{3}{c|}{$D=2$}
    \\
\cline{2-7}
 		& \multicolumn{3}{c|}{$E_0^{(1)}$}    &$\quad\quad\,E_0^{(1)}\quad\quad\,$       &$\quad\quad\,-E_2\quad\quad\,$         &$\quad\quad\,E_0^{(2)}\quad\quad\,$
    \\
\hline
 			\rule{0pt}{5ex}0.1                & \multicolumn{3}{c|}{5.436849553}    &6.528432582       &$2.43\times10^{-7}$        &6.52843233834
    \\
 		1.0                & \multicolumn{3}{c|}{7.879141644 }    &9.465319955        &$2.21\times10^{-8}$       &9.46531993256
    \\
 		10.0                 & \multicolumn{3}{c|}{16.641305904}   & 19.981458531       &$2.23\times10^{-7}$       &19.98145830814       \\[8pt]  \hline\hline
 		\multirow{2}{*}{$g$} & \multicolumn{3}{c|}{$D=3$} & \multicolumn{3}{c|}{$D=6$}
    \\
\cline{2-7}
 		& $\quad\quad\, E_0^{(1)} \quad\quad\,$ & $\quad\quad\, -E_2 \quad\quad\,$ & $\quad\quad\, E_0^{(2)} \quad\quad\,$ & $E_0^{(1)}$ & $-E_2$ & $E_0^{(2)}$
    \\
\hline
 	\rule{0pt}{5ex}
    0.1 & 7.652743974 & $1.87 \times 10^{-7}$ & 7.652743787 & 11.069434802 & $6.73 \times 10^{-8}$ & 11.069434735
    \\
 	1.0 & 11.224406591 & $1.87 \times 10^{-8}$ & 11.224406573 & 16.699837135 & $1.22 \times 10^{-8}$
   & 16.699837123
    \\
   10.0 & 23.860743313 & $1.78 \times 10^{-7}$ & 23.860743135 & 36.070426676 & $1.01 \times 10^{-7}$           & 36.070426576
    \\[8pt]
\hline
 	\end{tabular}
 \end{table}

\begin{table}[h]
	\caption{Variational energy $E_0^{(1)}$ of the third excited state with quantum numbers
      (1,0) - the first radial excitation for the potential  $(r^2 + g r^3)$ for $D=2,3,6$
      and at $g=0.1, 1, 10$ and its node $r_0^{(0)}$, found with use of $\Psi_{1,0}^{(t)}$, see text. Correction $E_2$ (not shown) contributes systematically to the 7th~d.d.}
      \label{cubicres3}
\resizebox{\textwidth}{!}{\begin{tabular}{|c|cc|cc|cc|}
\cline{1-7}
\multicolumn{1}{|c|}{\multirow{2}{*}{$\quad\,g\quad\,$}} & \multicolumn{2}{c|}{$D=2$} & \multicolumn{2}{c|}{$D=3$} & \multicolumn{2}{c|}{$D=6$}
   \\
\cline{2-7}
\multicolumn{1}{|c|}{} & $ \qquad \qquad E_0^{(1)}\ \ \ \ \ \ \ \ $ & $\ \ \ \ \ \ \ \ r_0^{(0)}\ \ \ \ \ \ \ \ $ & $\ \ \ \ \ \ \ \ E_0^{(1)}\ \ \ \ \ \ \ \ $ & $\ \ \ \ \ \ \ \ r_0^{(0)}\ \ \ \ \ \ \ \ $ & $\ \ \ \ \ \ \ \  E_0^{(1)}\ \ \ \ \ \ \ $ & $\ \ \ \ \ \ \ r_0^{(0)}\ \ \ \ \ \ \ \ \ $
    \\
\hline
	\rule{0pt}{5ex}
    0.1 & 6.570942086  & 0.953377788 & 7.709696613  & 1.162457356 & 11.15814973  & 1.626236134
    \\
    1.0 & 9.690374810  & 0.780305457 & 11.517370500 & 0.941956538 & 17.128462944 & 1.289494458
    \\
   10.0 & 20.681623429 & 0.532055331 & 24.758598615 & 0.638726047 & 37.346045552 & 0.865045854
    \\[8pt]
\hline
\end{tabular}}
\end{table}
The Non-Linearization Procedure allows us to estimate a deviation of the Approximant $\Psi_{n_r,l}^{(t)}$ from the exact wave function $\Psi_{n_r,l}$. As for the ground state the relative deviation is  bounded and very small,
\begin{equation}
\label{estimate}
 \left|\frac{\Psi_{0,0}(r)-\Psi_{0,0}^{(t)}}{\Psi_{0,0}^{(t)}}\right| \lesssim 10^{-4}
\end{equation}
in the whole range of $r \in [0, \infty)$ at any dimension $D$ we explored and at any coupling constant $g \geq0$ studied. Therefore, the Approximant leads to a locally accurate approximation of the exact wave function $\Psi_{0,0}(r)$ once the optimal parameters for $\Psi_{0,0}^{(t)}$ are chosen. These optimal parameters are smooth slow-changing functions in $D$ and $g$.
Similar situation appears for the excited states for different $D$ and $g$.

The Approximant $\Psi_{1,0}^{(t)}$ also provides an accurate estimate of the position of radial node of the exact wave function, see (\ref{trialcubic-psi-excited}). The orthogonality condition imposed to $\Psi_{0,0}^{(t)}$ and $\Psi_{1,0}^{(t)}$ provides a simple analytic expression for the zero order approximation $r_0^{(0)}$  to  $r_0$. Making comparison of $r_0^{(0)}$ with  numerical estimates which come from the Lagrange Mesh Method with 50 mesh points, we can see the coincidence of  $r_0$ and $r_0^{(0)}$ for at least 5 d.d. at integer dimension $D$ at any coupling constant $g \geq 0$. Results are presented in Table \ref{cubicres3}.

In all cases studied the first correction $y_1$ to the logarithmic derivative of the ground state function is not bounded, however, the ratio $|y_1/y_0|$ is bounded and small,
thus, $y_1$ is a \textit{small} function in comparison with $y_0$.
In Figs. 4 - \ref{fig:D=3} it is presented $y_0$ and $y_1$ {\it vs.} $r$ for $g=1$ in physical dimensions $D=1,2,3$. Similar plots appear for $D=6$. Making analysis of these plots one see that in domain $1 \gtrsim r \geq 0$, which gives dominant contribution to integrals defining the energy corrections, the $|y_1|$ is extremely small comparing to $|y_0|$ being close to zero.
It explains why the energy correction $E_2$ is small being the order of $\sim 10^{-7}$, or $\sim 10^{-8}$.
In similar way one can show numerically that the higher corrections $y_2, y_3, \ldots$ drop down to zero in the domain $1 \gtrsim r \geq 0$ even faster indicating the convergence both $y_n$ and $E^{(n)}$ as $n\rightarrow \infty$.

\begin{figure}[ht]
  	\includegraphics[width=0.99\textwidth]{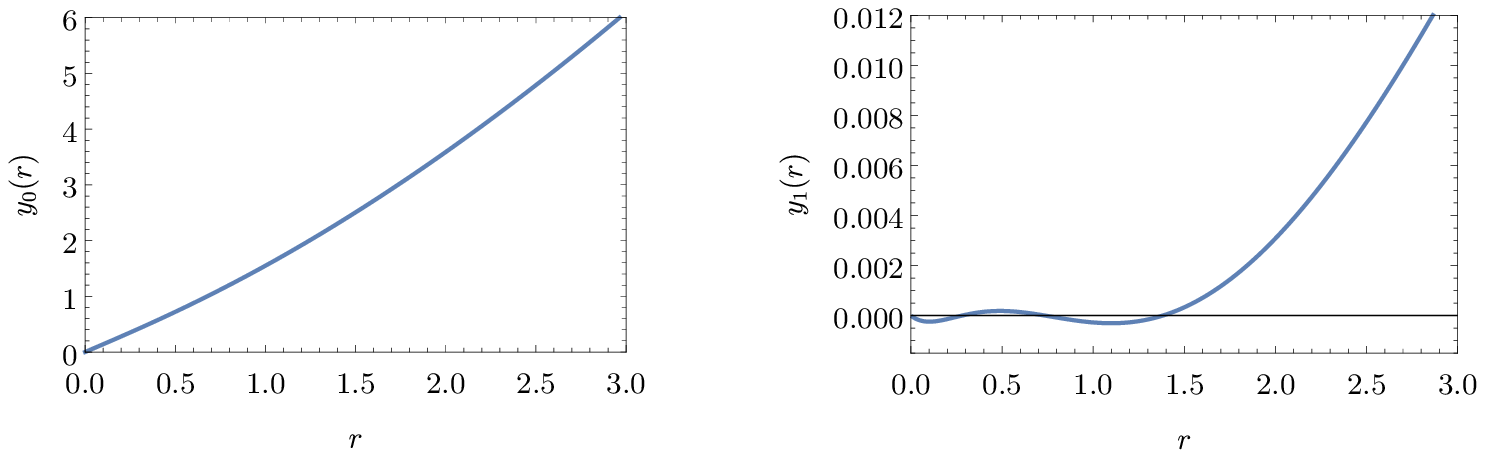}
  	\label{fig:D=1*}
  	\caption{Function $y_0=(\Phi_t)'$ (on left) and its first order correction $y_1$ (on right)
     as a function of $r$ at $D=1$ and $g=1$.}
\end{figure}
\begin{figure}[ht]
  	\includegraphics[width=0.99\textwidth]{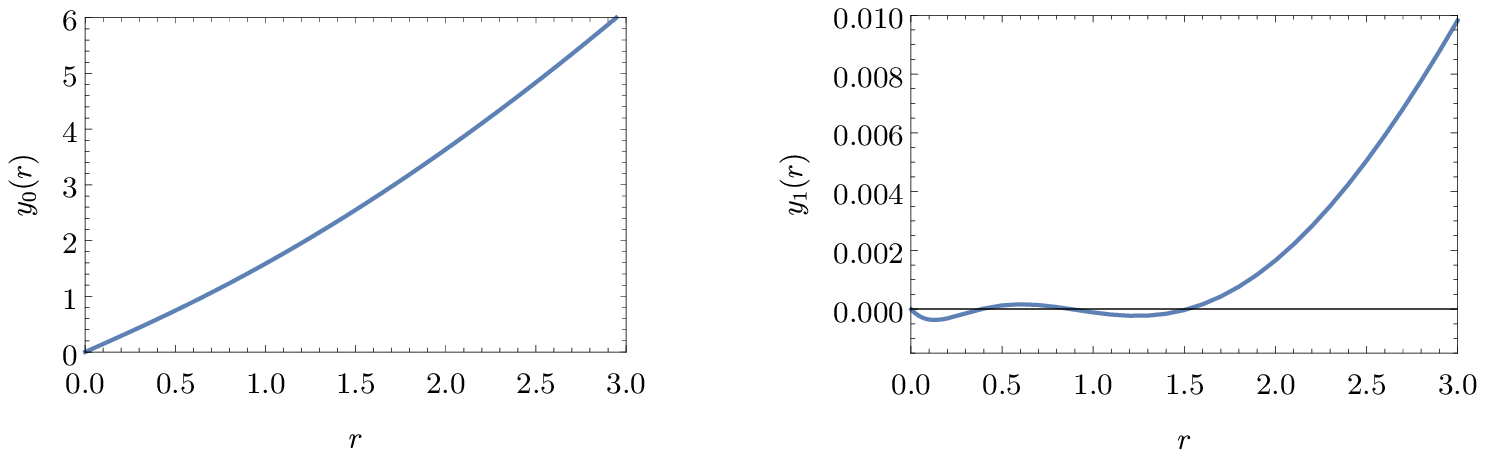}
  	\caption{Function $y_0=(\Phi_t)'$ (on left) and its first order correction $y_1$ (on right)
     as a function of $r$ at $D=2$ and $g=1$.}
  	\label{fig:D=2}
\end{figure}
\begin{figure}[ht]
  	\includegraphics[width=0.99\textwidth]{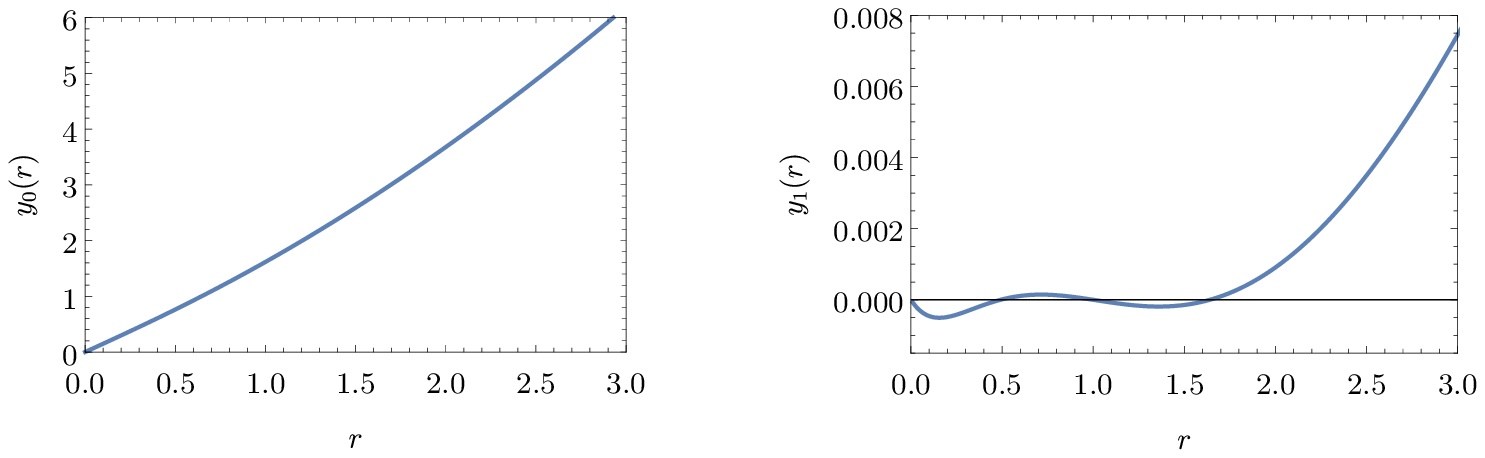}
  	\caption{Function $y_0=(\Phi_t)'$ (on left) and its first order correction $y_1$ (on right)
     as a function of $r$ at $D=3$ and $g=1$.}
  	\label{fig:D=3}
\end{figure}

\subsection{The Strong Coupling Expansion}

Here we present the results for the first two terms in the strong coupling expansion (\ref{energyST}) for the ground state energy of cubic oscillator,
\begin{equation}
\label{energySTcubic}
   E\ =\ g^{2/5}\left( {\tilde \veps}_0\ +\ {\tilde \veps}_1 g^{-4/5}\ +\
   {\tilde \veps}_2 g^{-8/5}\ +\ \ldots
   \right) \ ,
\end{equation}
assuming for simplicity $2M=\hbar=1$. It is worth mentioning that similar expansion holds for
any excited state. Evidently, it has finite radius of convergence.
This expansion corresponds to PT in powers of ${\hat \la}$ for potential
\begin{equation}
\label{potentialcubic-w}
       V(w)\ =\ w^3\ +\ {\hat \la}\, w^2\ ,\quad {\hat \la} \ =\ g^{-4/5}\ ,
\end{equation}
defined in $w \in [0, \infty)$.

As the first step we focus on calculation of ${\tilde \veps}_0$ in (\ref{energySTcubic}), which is, in fact, the ground state energy in the pure cubic radial potential $V = g r^3$ (ultra-strong coupling regime). In order to do it we begin with one of the simplest \textit{physically relevant trial function} (\ref{psi0}),
\begin{equation}
\label{eq:psi00}
 \Psi_{0,0}^{(t)}\ =\ e^{-\frac{2}{5}\,w^{5}}\ .
\end{equation}
for the ground state in potential (\ref{potentialcubic-w}) at ${\hat \la}=0$.
Taking $\Psi_{0,0}^{(t)}$ as zero approximation we calculate variational energy analytically and develop the PT procedure, described in Section \ref{strong}, numerically. In this manner we  obtain different PT corrections $\tilde{\veps}_{0,k}\,,\ k=0,1,2,3,\ldots$ to the exact value of $\tilde{\veps}_0$ and then form the partial sums (\ref{e01}).
The first non-trivial partial sum $(\tilde{\veps}_{0,k} + \tilde{\veps}_{0,1})$ is equal to variational energy.
Explicit results including partial sums up to sixth order are presented in Table \ref{table:stcubic1d} for the one-dimensional case, $D=1$. We have to note a slow convergence to the exact result with increase of the number of perturbative terms in expansion (\ref{e01}) taken into account. Even the partial sum $\sum_{k=0}^{6}\tilde{\veps}_{0,k}$ with seven terms included allows us to reproduce 7 s.d. {\it only} of the highly-accurate result obtained in Lagrange mesh method with 50 mesh points. It is natural to accelerate convergence taking more advanced trial function as entry. As for the second term in the expansion (\ref{energySTcubic}) it can be estimated by calculating with (\ref{eq:psi00}) the normalized expectation value of $w^2$,
\begin{equation}
\label{eps1}
    {\tilde \veps}_1 \ =\ \frac{ \langle\Psi_{0,0}^{(t)}|w^2|\Psi_{0,0}^{(t)}\rangle}{\langle\Psi_{0,0}^{(t)}|\Psi_{0,0}^{(t)}\rangle}\ =\ 0.495 \ .
\end{equation}
This straightforward estimate with simplest trial function (\ref{eq:psi00}) provides ${\tilde \veps}_1$ with accuracy $\sim 20\%$, see below. In similar way how it was done for ${\tilde \veps}_0$ one can develop PT for ${\tilde \veps}_1$ in (\ref{energySTcubic}). This PT is fast convergent: the first PT correction to (\ref{eps1}) is equal to $-0.086$: it leads to ${\tilde \veps}_1=0.409$  which differs from the exact value in $\sim 0.5\%$, see Table \ref{table:subdominantstrong}.

\begin{table}[ht]
\caption{Different partial sums (\ref{e01}) for the leading term $\tilde{\veps}_0$ of the strong coupling expansion (\ref{energySTcubic}) for the one-dimensional cubic anharmonic oscillator in PT with trial function (\ref{eq:psi00}) as zero approximation. Exact $\tilde{\veps}_0$ obtained via Lagrange mesh method with 50 mesh points.}
\label{table:stcubic1d}
\centering
\resizebox{\textwidth}{!}{\begin{tabular}{|c|c|c|c|c|c|c|c|}
\hline
 $\tilde{\veps}_{0,0}$ & $\sum_{k=0}^{1}\tilde{\veps}_{0,k}$ & $\sum_{k=0}^{2}\tilde{\veps}_{0,k}$ & $\sum_{k=0}^{3}\tilde{\veps}_{0,k}$ & $\sum_{k=0}^{4}\tilde{\veps}_{0,k}$ & $\sum_{k=0}^{5}\tilde{\veps}_{0,k}$ & $\sum_{k=0}^{6}\tilde{\veps}_{0,k}$ & Exact $\tilde{\veps}_0$
    \\
\hline
 \rule{0pt}{4ex} 0 & \quad\,1.053006976\quad\, & \quad\,1.021174929\quad\, & \quad\,1.022989568 \quad\, & \quad\,1.022956899 \quad\, & \quad\,1.022946414\quad\, &
           1.022947763 \quad\, & 1.022947875 \quad\,
    \\[4pt]
\hline
\end{tabular}}
\end{table}

As alternative to the trial function (\ref{eq:psi00}) let us use the Approximant (\ref{trialcubic}) in order to calculate the first two terms in the strong coupling expansion (\ref{energySTcubic}), see also (\ref{energyST}), for the ground state. In Table \ref{table:strong} we present for different $D$ the leading coefficient of the strong  coupling expansion $\tilde{\veps}_0$ found variationally $\tilde{\veps}_0^{(1)}$ and the second PT correction $\tilde{\veps}_2$ calculated to it via the Non-Linearization Procedure. We introduce the partial sum $\tilde{\veps}_0^{(2)}=\tilde{\veps}_0^{(1)}+\tilde{\veps}_2$.
The final results are verified using the Lagrange mesh method with 50 mesh points, see discussion above.
One can see that systematically the correction $\tilde{\veps}_2$ is of order
of $\sim 10^{-7}$. Hence, the first six decimal digits in variational energy are correct
and it defines the accuracy of variational calculations of ${\tilde \veps}_0$
with the Approximant (\ref{trialcubic}) as the trial function. One can estimate the order
of the third PT correction: all $D$ studied $\tilde{\veps}_3 \sim 10^{-2}\,\tilde{\veps}_2$. It indicates to extremely high rate of convergence
$\sim 10^{-2}$ of our PT. In Table  \ref{table:subdominantstrong} we present the first two approximations of the coefficient $\tilde{\veps}_1$ in (\ref{energySTcubic}).

\begin{table}[h]
\centering
\caption{Ground state energy $\tilde{\veps}_0$ (see (\ref{energySTcubic})) for the potential
         $W=r^3$ for $D=1,2,3,6$ found in PT with the Approximant $\Psi_{0,0}^{(t)}$ as the entry: $\tilde{\veps}_0^{(1)}$ is the variational energy,  $\tilde{\veps}_2$ is the second
         PT correction and $\tilde{\veps}_0^{(2)}=\tilde{\veps}_0^{(1)}+\tilde{\veps}_2$
         is corrected variational energy. Eight decimal digits in $\tilde{\veps}_0^{(2)}$
         confirmed in Lagrange mesh calculation.}
\label{table:strong}
\begin{tabular}{|ccc|ccc|}
\hline
			\multicolumn{3}{|c|}{$D=1$} & \multicolumn{3}{c|}{$D=2$}
    \\
\hline
			  $\quad\quad\,\tilde{\veps}_0^{(1)}\quad\quad\,$
            & $\quad\quad\,-\tilde{\veps}_2\quad\quad\,$
            & $\quad\quad\,\tilde{\veps}_0^{(2)}\quad\quad\,$
            & $\quad\quad\,\tilde{\veps}_0^{(1)}\quad\quad\,$
            & $\quad\quad\,-\tilde{\veps}_2\quad\quad\,$
            & $\quad\quad\,\tilde{\veps}_0^{(2)}\quad\quad\,$
    \\
\hline
			\rule{0pt}{4ex}
      1.022948250 & $3.75 \times 10^{-7}$ & 1.022947875 & 2.187461809 & $3.09 \times 10^{-7}$ & 2.187461499
    \\[4pt]
\hline
\hline
			\multicolumn{3}{|c|}{$D=3$} & \multicolumn{3}{c|}{$D=6$}
    \\
\hline
			$\tilde{\veps}_0^{(1)}$ & $-\tilde{\veps}_2$ & $\tilde{\veps}_0^{(2)}$ & $\tilde{\veps}_0^{(1)}$      & $-\tilde{\veps}_2$ & $\tilde{\veps}_0^{(2)}$
    \\
\hline
			\rule{0pt}{4ex}
      3.450562918 & $2.29 \times 10^{-7}$ & 3.450562689 &
            7.647118254 & $1.01 \times 10^{-7}$ & 7.647118153
    \\[4pt]
\hline
		\end{tabular}
\end{table}
	
\begin{table}[h]
		\centering
		\caption{Subdominant (next-to-leading) term $\tilde{\veps}_1$ in the
         strong coupling expansion (\ref{energySTcubic}) of the ground state energy for the cubic anharmonic radial
         potential for $D=1,2,3,6$.}
\label{table:subdominantstrong}
\begin{tabular}{|ccc|ccc|}
\hline
			\multicolumn{3}{|c|}{$D=1$} & \multicolumn{3}{|c|}{$D=2$} \\
\hline
	$\quad\quad\,\tilde{\veps}_1^{(1)} \quad\quad\,$     &
    $\quad\quad\,\tilde{\veps}_{1,1}   \quad\quad\,$     &
    $\quad\quad\,\tilde{\veps}_1^{(2)} \quad\quad\,$     &
    $\quad\quad\,\tilde{\veps}_1^{(1)} \quad\quad\,$     &
    $\quad\quad\,\tilde{\veps}_{1,1}   \quad\quad\,$     &
    $\quad\quad\,\tilde{\veps}_1^{(2)} \quad\quad\,$     \\
\hline
		\rule{0pt}{4ex}0.410598524 & $6.78\times10^{-7}$ & 0.410599202 &
         0.766573847  & $5.24\times10^{-7}$ & 0.766574371  \\[4pt]
\hline
\hline
		  \multicolumn{3}{|c|}{$D=3$} & \multicolumn{3}{|c|}{$D=6$}   \\
\hline
		$\tilde{\veps}_1^{(1)}$ & $\tilde{\veps}_{1,1}$    &
        $\tilde{\veps}_1^{(2)}$ & $\tilde{\veps}_1^{(1)}$  & $\tilde{\veps}_{1,1}$   & $\tilde{\veps}_1^{(2)}$  \\
\hline
		\rule{0pt}{4ex}1.092125224 & $2.67\times10^{-7}$ & 1.092125491 & 1.967599668
        & $1.42\times10^{-7}$      & 1.967599810            \\[4pt]
\hline
\end{tabular}
\end{table}

Summarizing the studies of $D$-dimensional radial cubic anharmonic oscillator for a given $D$, one can imagine that the numerical results for given eigenvalue can be \textit{described} via a simple analytical formula. Inspiration comes from the fact that the eigenvalue {\it vs} coupling constant $g$ is very smooth, slow growing function.
Basic idea is to interpolate the expansions in $g$ for the weak and strong coupling regimes, see (\ref{seriespE}) and (\ref{energySTB}). It has been already shown that this approach is appropriate and successful in description of various physical systems of different nature, e.g.  \cite{delValle} and  \cite{OLIVARESPILON}. In particular, for the ground state of the cubic radial anharmonic potential, the simplest interpolations is of the form
\begin{equation}
    E(g)\ \sim\  D(1 + a\, g+ b^5\,g^2)^{\frac{1}{5}}\ ,
\label{int_energy}
\end{equation}
where $a$ is a free parameter; we set $b = (\tilde{\veps}_0/D)$, it is slow-changing parameter with $D$, see Table \ref{table:strong}. This interpolation reproduces exactly both leading terms in (\ref{seriespE}) and (\ref{energySTcubic}). Parameter $a$ is fixed by fit of numerical data with minimal $\chi^2$. Optimal parameter $a$ and the parameter $b$ are presented in Table \ref{inter}. For integer $D$ the above simple formula describes the ground state energy $E$ for any $g \in [0, \infty)$ with accuracy $\lesssim 2 \%$.
In a similar manner a straightforwardly modified formula works for energies of excited states.
\begin{table}[h]
\caption{Parameters of fit (\ref{int_energy}) for different $D$.}
\label{inter}
\begin{tabular}{|c|cccc|}
\hline
	 Parameter&  $\quad D=1$\quad &  $\quad D=2\quad $&  $\quad D=3\quad $&$\quad D=6\quad$  \\
\hline
		\rule{0pt}{4ex}$a$ & 3.281  & 3.922  & 4.823 & 5.994  \\
	               $b$ & 1.023 & 1.094  & 1.150  & 1.275 \\[4pt]
\hline
\end{tabular}
\end{table}
	
For one-dimensional quartic oscillator $V=x^2+g x^4$ it was discovered long ago
by Bender and Wu \cite{BENDER} and recently proved rigorously by Eremenko and Gabrielov \cite{Eremenko:2009} that even (odd) parity eigenvalues form infinitely-sheeted Riemann surface in space of coupling constant $g$. This surface is characterized by infinitely many square-root branch points having a meaning of the points of level crossings at complex $g$.
Each Riemann sheet contains infinitely many such square-root branch points manifesting
that every two energy levels intersect. Seemingly, the same phenomenon holds for any one-dimensional anharmonic oscillator with two term potential $V=x^2+g x^{2m}$ defined in the whole line $x \in (-\infty, +\infty)$. We guess that for cubic anharmonic oscillator
$V=r^2+g r^3$ at $r \in [0, \infty)$ for fixed integer dimension $D > 1$ and each value of the angular momentum $\ell=0,1,\ldots$ the energies with quantum numbers $(n_r,\ell), n_r=0,1,2,\ldots$ form infinitely-sheeted Riemann surface in variable ${\tilde \la} = g^{-4/5}$
with square-root branch points. It is evident that due to the Hermiticity of the original Hamiltonian (1) each branch point should have its complex-conjugate. It could be a definite challenge to calculate the distance ${\tilde \la}^{\star}$ from the origin to the nearest square-root branch point in ${\tilde \la}$ and find its $D$ dependence, since it defines the radius of convergence of strong coupling expansion for two intersecting energy levels $(n_r,\ell)$ and $({\tilde n_r},\ell)$.
The most important particular case is to find ${\tilde \la}^{\star}$ for the ground state $(0,0)$ and the first excited state $(1,0)$: it probably defines the radius of convergence
of the strong coupling expansion (\ref{energySTcubic}). It is not clear what intersection of the ground state eigenvalue with one of the excited state is closest to the origin in ${\tilde \la}$. This challenge will be tackled elsewhere.

\section*{Conclusions}

There are two views of anharmonic, polynomial, radial oscillator potential:
(i) as in coordinate $r$-space it is a finite degree perturbation in coupling constant $g$ being polynomial in $r$ perturbation of the harmonic oscillator and
(ii) as finite-degree polynomial in {\it classical} coordinate $(g r)$ with coupling constant dependence $1/g^2$ in front of the potential. These views suggest to make study the dynamics in these two spaces: $r$-space and in $(gr)$-space. It leads to two different non-linear differential equations for the logarithmic derivative of the wavefunction, both are of the first order with quadratic non-linearity: the RG equation and the GB equation with the {\it same} effective coupling constant $\la$ and energy $\veps$.
Perturbation theory in powers of $\la$ developed in RG equation leads effectively to a description of small $r$ domain, while developed in GR equation leads effectively to a description of large $(gr)$ domain. Latter generates a new version of semiclassical expansion in fractional powers of $\hbar$ in a very compact way. In complementary way the perturbation theory in inverse fractional powers of $\la$ developed in RG equation leads effectively to a description of large $r$ domain, while developed in GR equation leads effectively to a description of small $(gr)$ domain. Latter generates {\it anti}-semiclassical expansion in inverse fractional powers of $\hbar$.

Combining all four expansions together, we are able to construct for logarithmic derivative $y$
(\ref{riccati1d}) the interpolating function $y_0$ (\ref{generalrecipe}) for both the general case (\ref{potential}) and the particular case (\ref{trialcubic}) of cubic oscillator (\ref{cubicpotential}).
Eventually, we arrive to 3-parametric trial function for any eigenstate of cubic oscillator.
It seems evident that, in general, $y_0$ should be very close to the exact $y$ -
the first correction $y_1$ is small, $|y_1/y_0| \ll 1$ for any $r \in [0, \infty)$. We show that by
taking $y_0$ as zero approximation the Non-Linearization procedure appears fast convergent. It is explicitly checked for cubic oscillator that the relative deviation of the Approximant (\ref{trialcubic-psi}) from exact ground state eigenfunction is less than $10^{-4}$, see (\ref{estimate}), for any $r \in [0, \infty)$. Simultaneously, the absolute accuracy in energy reaches the extremely high value $\sim 10^{-7}$ at any dimension $D$ and coupling constant $g \in [0, \infty)$. Similar accuracy can be reached for any expectation value for which the dominant contribution to integrals comes from the domain $r \in [0, 1]$.
It manifests a solution of the problem of ground state of cubic radial anharmonic oscillator as well as its low-lying excited states! Based on this formalism similar accuracy is reached for quartic and sextic radial anharmonic oscillators that will be presented in subsequent paper \cite{delValle2}.

All studies performed lead to smooth dependence on dimension $D$. We did not see any indication that physics dimension $D=3$ is special.

\section*{Acknowledgments}
The authors thank J.C.~L\'opez~Vieyra and H.~Olivares~Pil\'on for their interest to the work and useful remarks, and especially for help with numerical and computer calculations. J.C.~del\,V. is supported by a CONACyT PhD Grant No.570617 (Mexico). A.V.T. thanks the Stony Brook University and Simons Center for Geometry and Physics at Stony Brook, NY, and especially, E Shuryak for kind hospitality extended to him, where some of present work was carried out.  This work is partially supported by CONACyT grant A1-S-17364 and DGAPA grant IN108815 (Mexico).


\appendix

\section{First PT Corrections and Generating Functions $G_{3,4}$ for the Cubic Anharmonic Oscillator}

\label{appendix:A}

The expansions in powers of $v$ for the first three corrections $\mathcal{ Y}_n(v), n=1,2,3$
for the cubic anharmonic oscillator potential (\ref{cubicpotential}), see (\ref{small}) and (\ref{large}) are presented.

\begin{itemize}
\item At $v\rightarrow0$
	\begin{align}
		\mathcal{ Y}_1(v)&\ =\ \epsilon_1v\ +\ \frac{ 2\, \epsilon _1}{D+2}v^3\ -\ \frac{1}{D+3}v^4\ +\ \frac{ \epsilon _1}{(D+2) (D+4)}v^5\ +\ \ldots\ , \non\\
		\mathcal{ Y}_2(v)&\ =\ \epsilon_2\,v\ +\ \frac{ \epsilon _1^2\ +\ 2\, \epsilon_2}{D+2}v^3\ +\ \frac{6\, \epsilon _1^2+4\,  \epsilon _2}{ (D+2) (D+4)}v^5\ -\ \frac{2\,\epsilon _1}{(D+3) (D+5)}v^6\ +\ \ldots\ , \non\\
		\mathcal{ Y}_3(v)&\ =\ \epsilon_3\,v +\frac{2 (\epsilon _1\, \epsilon _2+ \epsilon _3)}{ D+2}v^3\ +\ \frac{2 (\epsilon _1^3+6\,  \epsilon _1\, \epsilon _2+2\,  \epsilon _3)}{ (D+2) (D+4)}v^5\ -\ \frac{2\, \epsilon _2}{ (D+3) (D+5)}v^6\ +\ \ldots\ .
	\end{align}
In these formulas we have denoted  $\epsilon_n=\frac{\veps_n}{D}$.
	
\item At $v \rar \infty$
	\begin{align}
		\mathcal{Y}_1(v)\ =\ &\frac{1}{2}v^2\ +\ \frac{1}{4}(D+1)\ -\ \frac{\veps _1}{2}v^{-1}\ +\ \frac{1}{8} (D^2-1)v^{-2}\ +\ \ldots\ ,\non\\
			\mathcal{Y}_2(v)\ =\ &-\frac{1}{8}v^3-\frac{1}{16} (3 D+4) v\ +\
 \frac{\veps _1}{4}\ -\ \frac{1}{32 }(6 D^2+6 D+16\, \veps _2-1)v^{-1}\non\\
		&+\frac{1}{8} (3 D-2) \,\veps _1 v^{-2}\ +\ \ldots\ ,\nonumber\\
			\mathcal{Y}_3(v)\ =\ &\frac{1}{16}v^4\ +\ \frac{1}{32} (5 D+8) v^2\ -\ \frac{3\,\veps _1}{16}  v\ +\ \frac{1}{64} (15 D^2+26 D+16\, \veps _2+10)\nonumber\\
		&-\frac{1}{32}(15 D\, \veps _1+16\, \veps _3)v^{-1}\ +\ \ldots\ .
	\end{align}
\end{itemize}

A straightforward application of formulas (\ref{BlochSol}) and (\ref{semiclassicalcorrection}) allows us to calculate the generating functions. In particular, $G_3$ and $G_4$ are given by
\begin{align}
	G_3\ =\ &-\frac{(2M)^{1/2}}{g^2}
     \left(\frac{\veps _1}{2} \log\left[\frac{w-1}{w+1}\right]\right)\ ,\non\\
	G_4\ =\ &-\frac{(2M)^{1/2}}{g^2} \left(\frac{5 + (5 + 12 D) w+(1-6 D(D+1)) (w+1)w^2
     \left(3 w^2-2\right)}{48  (w-1) (w+1)^2w^3 } \right.\non\\
	& \left. \frac{\left(1-16\,\veps_2 -6 D(D+1)\right)}{32 }
     \log \left[\frac{w-1}{w+1}\right]\right)\ ,
\end{align}
where $w=(1+gr)^{1/2}$.

\bibliography{references1}

\end{document}